\documentclass[12pt]{article}

\usepackage{setspace}
\usepackage{latexsym}
\usepackage{amsmath}
\usepackage{amssymb}
\usepackage{relsize}
\usepackage{geometry}
\usepackage{amsmath, amssymb, amscd, xypic, graphicx}
\usepackage{slashed,color}
\usepackage{epstopdf}
\usepackage[colorlinks]{hyperref}
\usepackage[colorinlistoftodos, textwidth=4cm, shadow]{todonotes}
\usepackage[displaymath, tightpage]{preview}
\usepackage{titlesec}
\usepackage{hyperref}
\usepackage{cite}

\input xy
\xyoption{all}
\numberwithin{equation}{section}
\numberwithin{equation}{subsection}
\def\beq{\begin{equation}}
\def\eeq{\end{equation}}
\def\beqn{\begin{eqnarray}}
\def\eeqn{\end{eqnarray}}
\def\Tr{{\rm Tr}}
\def\T{{\rm T}}

\def\mf{\mathfrak}
\def\dx{{\rm d}^2x}

\newcommand{\cw}{{\mathcal W}}

\newcommand{\sma}{\left(\begin{smallmatrix}}
\newcommand{\smaa}{\end{smallmatrix}\right)}

\newcommand{\gsim}{\lower.7ex\hbox{$
\;\stackrel{\textstyle>}{\sim}\;$}}
\newcommand{\lsim}{\lower.7ex\hbox{$
\;\stackrel{\textstyle<}{\sim}\;$}}

\begin{document}

\vspace{0.8cm}

\begin{titlepage}

\begin{flushright}
FTPI-MINN-15/42,\\
UMN-TH-3503/15~~\\
\end{flushright}

\vspace{1cm}

\begin{center}
{  \Large \bf  Anomalies of Minimal \boldmath{$\mathcal N=(0, 1)$ and $\mathcal N=(0, 2)$}\\[2mm] Sigma Models on Homogeneous Spaces}

\vspace{0.8cm}

{\large
Jin Chen,$^{a}$ Xiaoyi Cui,$^{b}$ Mikhail Shifman,$^{a,c}$ \\[2mm] and Arkady Vainshtein$^{\,a,c}$}
\end {center}

\vspace{1mm}

\begin{center}

$^{a}${\it School of Physics and Astronomy, University of Minnesota,\\
Minneapolis, MN 55455, USA}\\[1mm]
$^{b}${\it Mathematisches Institut, Georg-August Universit\"{a}t G\"{o}ttingen,\\ G\"{o}ttingen, D-37073, Germany}\\[1mm]
$^c${\it  William I. Fine Theoretical Physics Institute,
University of Minnesota,
Minneapolis, MN 55455, USA}\\[1mm]

\end{center}

\vspace{0.5cm}

\begin{center}
{\large\bf Abstract}
\end{center}

We study chiral anomalies in $\mathcal N=(0, 1)$ and $(0, 2)$ two-dimensional minimal sigma models defined on generic homogeneous spaces $G/H$. Such minimal theories contain only (left) chiral fermions and in certain cases
are  inconsistent because of ``incurable" anomalies. We explicitly calculate the anomalous fermionic effective action and show how to remedy it by adding a series of local counterterms. In this procedure, we derive a local anomaly matching condition, which is demonstrated to be equivalent to the well-known global topological constraint 
on $p_1(G/H)$, the first Pontryagin class. More importantly, we show that these local counterterms further modify and constrain  ``curable" 
chiral models,  some of which, for example, flow to nontrivial infrared superconformal fixed point. Finally,  we also observe an interesting relation between $\mathcal N\!=\!(0, 1)$ and $(0, 2)$ two-dimensional minimal sigma models and supersymmetric gauge theories.

This paper  generalizes and extends the results of our previous publication arXiv: 1510.04324.

\hspace{0.3cm}

\end{titlepage}

\tableofcontents

\newpage


\section{Introduction and summary}

Supersymmetric nonlinear sigma models in two dimensions present a useful theoretical laboratory which has deep connections with many other quantum theories, as well as with aspects of topology. In supersymmetric theories
rather often simplicity of the theory increases with the number of supercharges. 
 By simplicity we mean that the theory under consideration can have special properties allowing one to obtain exact results or uncover elegant mathematical structures. On the other hand, theories with less supersymmetry,
 presenting more difficulties for theoretical analysis, are sometimes closer to 
 physical phenomena, and as such must be thoroughly studied. 
In this paper, we will focus on minimal supersymmetric models  with $\mathcal N=(0,1)$ or $\mathcal N=(0,2)$ supersymmetry \cite{1, Gates:1987, Witten:2005px, bai2, bai3, Cui:2011rz, CCSV2}. It has long been known that such models, generally speaking,  exhibit chiral fermion anomaly which imposes severe constraints on the topology of the target manifold \cite{Witten:2005px, Moore:1984ws, Freed}. Due to this reason, such minimal supersymmetric models  are explored to a lesser extent than non-chiral models. The guiding principle established  \cite{Moore:1984ws} for the chiral  $\mathcal N=(0,1)$ or $\mathcal N=(0,2)$ sigma models is the first Pontryagin class.

Our present work is motivated by the following consideration. Firstly, the global anomaly cancellation condition does not touch the local behavior of the theory. Even when one has a ``good" theory, which has no global anomaly \cite{Moore:1984ws}, it does not automatically mean that one gets the well-defined theory for free. Ease of global anomaly only implies that one is able to introduce ``local counterterms" to correctly integrated out chiral fermions and find the anomaly-free fermionic effective action. The ``local counterterm" here is not to be confused with the terms added to absorb various divergences in the process of renormalization, since the quantization of fermions in two dimensions is insensitive to RG flow. In fact the roles played by what we call local counterterms are similar to that of the contact term in gauge theories, which is added to keep the transversality of certain polarization operators. Since the latter is sometime also referring to the Schwinger term, we refrain from using it here. Moreover, by explicitly curing such a theory (i.e. adding appropriate local counterterms), one can exhibit many quantum aspects of the theory in a more understandable way, thus enabling one to initiate a discussion of the 
infrared (IR) behavior of the theory, which was not carried out previously. 

Secondly, many sigma models have more than one equivalent formulations: a nonlinear description based on the Riemannian metric that encodes geometric information, an embedding into a larger linear target space and then imposing extra gauge symmetries, or constraints, or  a hybrid way lying between the above two formulations \cite{CCSV2}. Although classically all these formulations are equivalent, at the
quantum level one could have different considerations depending on the formulation. For example, in the nonlinear formulation it is easy to understand the global chiral fermion anomaly, while using the gauge formulation, one will be focused on the gauge anomaly.  Work has been done on these aspects \cite{Moore:1984ws, Freed, Zumino:1983rz, AlvarezGaume:1983cs, AlvarezGaume:1984dr, Manohar:1984uq, Bagger:1985,  Manohar:1984zj, AlvarezGaume:1985yb}, providing us with starting positions. The precise relation between the gauge anomaly and global anomaly for different formulations of the very same model was not thoroughly discussed previously. In this paper, we  study of the chiral sigma models on homogeneous spaces, for which both nonlinear and gauge formulations are present. We  reveal the relation between different anomalies. Our result also provides us with a generalized context  for the determinant line bundle consideration of the fermion anomaly. In non-homogeneous spaces one can not compare global and gauge anomalies on the nose, but an analogous structure was revealed in the case of the K\"ahler manifold. This paper  generalizes and extends the results of \cite{CCSV2}.

Finally, we would like to emphasize  possible applications of our results in model building. Models with large supersymmetry may be viewed as being composed of theories with less supersymmetry. In this regard, understanding
of the minimal supersymmetric models as the building blocks for all supersymmetric theories is of importance. 

In practice, the usual situation is opposite.  For example, $\mathcal N = (2,2)$ theories are always better understood than $\mathcal N = (0,2)$ and explored earlier. Softly broken to $\mathcal N=(0,2)$ theories (free from chiral anomalies) they are easier for explorations \cite{Edalati:2007vk, Shifman:2008wv, Shifman:2008kj, B1, Cui:2010si, Cui:2011uw, Shifman:2014jba, Jia, Gadde:2013lxa, Gadde:2014ppa, CCSV1}. We hope that our work on the minimal models can give insights for understanding of more complicated models. 

We discuss at length the IR behavior for many models and observe a new connection with superconformal models \cite{Adams:2003zy, Melnikov:2012hk, Gadde:2014ppa}.

The paper is organized as follows. In Sec.~\ref{IA}, we will first  construct bosonic and $\mathcal N=(0, 1)$ supersymmetric sigma models on homogeneous spaces by virtue of a hidden gauge formulation. The calculation of their isometry anomalies is given in Sec.~\ref{2.3}. As discussed in the previous work \cite{CCSV2}, we show that the isometry anomalies reflect the failure of bundle re-parameterization from local section $s$ to $s^\prime$ induced by the isometry transformations, where $s,\ s^\prime: U_s\cap U_{s^\prime}\subset M\rightarrow G$. 

 To offset these aforementioned anomalies, we are led to consider (Sec.~\ref{HA}) more generic holonomy anomalies, of which isometry anomalies are a special class. We give criteria ensuring holonomy as well as isometry anomalies to be removed by adding well-defined local counterterms  (Sec.~\ref{3.1}). With these  criteria, and after adding
 appropriate local counterterms, we discuss the low-energy behavior of the minimal $\mathcal N=(0, 1)$ sigma models (Sec.~\ref{3.2}). In Sec.~\ref{3.3} several concrete examples are given to illustrate the idea. We review  appropriate tools that we had developed before. The topological origin of the anomalies and counterterms are discussed in Sec.\ref{3.4}. 

In Sec.~\ref{4} we begin to relate the holonomy and isometry anomalies to topological anomalies in a general context. In Sec.~\ref{4.1} a discussion of the isometry anomalies in the general K\"ahler sigma models is given, parallel to the relation between the non-Abelian gauge anomaly and chiral anomaly in gauge theories. In Sec.~\ref{4.2} we show how the isometry anomaly in pure geometric formulation relates to the topological chiral fermion anomaly in terms of the determinant line bundle discussed by Moore and Nelson, and Freed. In Sec.~\ref{4.3} we give the determinant line bundle description for the holonomy anomaly for sigma models over homogeneous spaces.  This completes a unified picture showing  that the holonomy (gauge) anomaly and the topological anomaly are due to the nontriviality of a single determinant line bundle over the space of fields.  

\section{Isometry anomalies}
\label{IA}
We will formulate this section by following the logic line of our previous work \cite{CCSV2} where we construct sigma models on $S^{2N-1}$ and gauge its $\rm U(1)$ factor to deduce the corresponding ${\rm CP}^{N-1}$ models by the fiberation:
$$\xymatrix{{\rm U}(1) \ar[r]^i & \,{\rm S}^{2N-1}\, \ar[r]^{\pi} &\,{\rm CP}^{N-1}}.$$
Similarly for homogeneous spaces, we also have a canonical fiberation:
$$\xymatrix{H \ar[r]^i & G \ar[r]^{\pi} & M}.$$
Therefore we first construct sigma models on group manifold $G$, and gauge certain subgroup $H$ to obtain sigma model on homogeneous spaces $M$. Analogue to ${\rm CP}^{N-1}$ case, to define a sigma model on $M$, one needs to specify a local patch $U_s\subset M$ and a section $s: U_s\rightarrow G$. To discuss isometry anomalies on model $M$, we will show that an isometric transformation $l_k: M\rightarrow M$ will induce a change of section $s$ to $s^\prime$, and thus, a $H$-gauge transformation. For chiral fermions non-trivially coupled to these $H$-gauge, there will be isometry anomalies produced. We will calculate them by the end of this section. For simplicity, we only consider $G$ as a connected, compact and semi-simple Lie group and $H$ is its closed Lie subgroup.

\subsection{Sigma models on \boldmath{$M$} through gauge formulation}

For sigma model on $M$, the construction can be traced back to 70's due to Callan-Coleman-Wess-Zumino (CCWZ) coset construction \cite{CCWZ}. In this subsection, we will review this construction but from the so-called ``hidden" local gauge formulation, which will be eventually explained in the language of principal bundle.

To have sigma model on $M$, as mentioned in the beginning, one first construct sigma model on group manifold $G$, and then ``gauge" it down to that of space $M$. We will see soon that such a construction is just the formulation with a ``hidden" local right-$H$ gauge, see \cite{Bando:1988}, in which the Nambu-Goldstone bosons are taking values in the group $G$ instead of $M$, and the right local $H$-gauge help eliminate redundant degree of freedom. Each time that one chooses a fixed gauge is equivalent to choose a local section to ``pullback" the model defined on bundle $G$ to base space $M$, and thus the language of principal bundle will be an ideal mathematical framework to interpret the model and further anomalies if there are any.\\

Since $G$ is semi-simple, one can always use the Killing form $K: \mf{g}\times\mf{g}\rightarrow \mathbb R$, which is negative definite, to define the metric $\,\bar{\!\gamma}$ of $G$. We consider the Lie algebra $\mf g$ in its fundamental representation,\footnote{\,It is true that Killing form is defined by means of adjoint representation of $G$, but for semi-simple Lie algebra one is free to rescale a constant for each simple factor and thus we can choose fundamental representation as our bench mark.} and normalize the anti-hermitian generators $F_A$ as:
\beq
K(F_A, F_B)=\Tr(F_AF_B)=-\delta_{AB}\ .
\label{Killing}
\eeq
In most of the note, we focus on sigma models defined on simple groups $G$. For bosonic sigma model on such a group $G$, the action is given by
\beq
S_G=\frac{1}{2\lambda^2}\int_\Sigma\dx\,{\rm Tr}(\partial_\mu g^{-1}\partial^\mu g)=-\frac{1}{2\lambda^2}\int_\Sigma\dx\,{\rm Tr}(g^{-1}\partial_\mu gg^{-1}\partial^\mu g),
\label{L_G}
\eeq
where $\Sigma$ is the two-dimensional spacetime manifold, $g=g(x)$ taking value on matrix group $G$, and $\lambda^2$ is a coupling constant.\footnote{\,For a semi-simple Lie group $G$, there are as many coupling constants $\lambda^2_i$ as the number of its simple factors $G_i$, and the Killing form $K$ is the direct sum of $K_i$ for each $G_i$.} It is seen in Eq.\,(\ref{L_G}) that $g^{-1}\partial^\mu g$ is the Maurer-Cartan form $\theta_g\equiv g^{-1}dg$ pullback to the cotangent space of spacetime $\Sigma$. For $g^{-1}dg\in\T^\ast G$ on $G$ defines map:
\beq
\theta_g=L_{g^{-1}\ast}: \T_g G\rightarrow\T_eG=\mathfrak{g}\,,
\label{mc1formglobal}
\eeq
where $L_{g^{-1}\ast}$ is the pushforward map induced by left translation $L_{g^{-1}}$, and $\T_e G$ is the tangent space of $G$ at group identity $e$, we thereby have the metric $\,\bar{\!\gamma}$ defined as
\beqn
\bar{\!\gamma}\,(X_g, Y_g)\equiv-K(L_{g^{-1}\ast} X_g, L_{g^{-1}\ast} Y_g)=-L_{g^{-1}}^\ast K(X_g, Y_g)\ ,
\label{metricG}
\eeqn
where $X_g$ and $Y_g$ are two vector fields at point $g\in G$. 

On a local chart $\{U, \phi^\alpha\}$ near identity $e\in G$, we can use exponential map to express $g(x)$ 
as\,\footnote{\,We use Greek and capital letters to distinguish the indexes of curved coordinates and that of flat vector space $\mathfrak{g}$}
$$g(x)={\rm Exp}(\delta_\alpha^A\phi^\alpha(x) F_A),\ \ {\rm for}\ A, \alpha=1,2,...,{\rm dim}\ G\,,$$
where $\phi^\alpha(x)$ are Nambu-Goldstone bosons. Therefore one can express $\theta_g$ and $\bar{\!\gamma}$ in a more familiar way as
\beq
\begin{split}
&\theta(\phi)=\theta^A_{\ \alpha}(\phi)d\phi^\alpha F_A\,,\\[1mm]
&\bar{\gamma}_{\alpha\beta}(\phi)=\delta_{AB}\theta^A_{\ \alpha}\theta^B_{\ \beta}\,,
\end{split}
\label{mc1form}
\eeq
where $\theta^A_{\ \alpha}$ is the vielbein to decompose $\bar{\gamma}_{\alpha\beta}$.
Notice that the vielbein one-form is left invariant, and right equivariant,
\beq
\begin{split}
&L_{g_0}^\ast\theta=(g_0g)^{-1}d(g_0g)=\theta\\[1mm]
&R_{g_0}^\ast\theta=(gg_0)^{-1}d(gg_0)=g_0^{-1}\theta g_0,\ ~~~{\rm for}\ g_0\in G\,.
\end{split}
\eeq
The metric $\bar{\!\gamma}$ defined above is consequently left and right invariant,
$$L_{g_0}^\ast\bar{\!\gamma}=R_{g_0}^\ast\bar{\!\gamma}=\bar{\gamma},\ {\rm for\ any}\ g_0\in G\,.$$
Therefore, the action $S_G$ has isometries $G_L\times G_R$\,.\\\\

\noindent Now we consider group $G$ as a principal bundle with fiber $H$ and base space \mbox{$M\!\equiv\! G/H$,}
$$\xymatrix{H \ar[r]^i & G \ar[r]^\pi & M}$$
with the projection
\beq
\begin{split}
&\xymatrix{\pi: G \ar[r] & M}\\[1mm]
&\hspace{6mm} \xymatrix{g\ar@{|->}[r]& gH}
\end{split}
\label{Hbundle}
\eeq
and $H$-group action acting from right on $G$, satisfies $\pi(gh)=\pi(g)$.

To define a sigma model on $M$, we notice formula (\ref{Hbundle}) that $H$-group action is from right to obtain $M$ coset space. It motivates us to gauge part of right isometries $H\subset G_R$ of sigma model on group $G$. Consider, $g(x)\rightarrow g(x)h(x)$ for a right $h(x)\in H$ transformation, the Maurer-Cartan form changes as:
\beq
g^{-1}dg\rightarrow h^{-1}(g^{-1}dg)h+h^{-1}dh\,.
\label{hgaugeformc}
\eeq
To make it gauge invariant, we introduce gauge fields
\beq
A(x)=A^i_{\mu}(x)dx^\mu H_i \,,
\label{hidengauge}
\eeq
where $H_i\in \mathfrak{h}\ {\rm for}\ i=1,2,...,{\rm dim} \mathfrak{h}$, taking values on Lie subalgebra $\mathfrak{h}$. It transforms as
$$A\rightarrow h^{-1}A h+h^{-1}dh$$
to remedy the additional $h^{-1}dh$ part of gauge transformation of $g^{-1}dg$. Therefore 
$$g^{-1}dg-A\rightarrow h^{-1}(g^{-1}dg-A)h$$
is gauge covariant. The action on $M$ is thus given by
\beq
S_{M}=-\frac{1}{2\lambda^2}\int_\Sigma\dx\,{\rm Tr}\big[(g^{-1}\partial_\mu g-A_\mu)(g^{-1}\partial^\mu g-A^\mu)\big]\ .
\label{L_G/H}
\eeq
After an appropriate gauge fixing, the action above will give usual CCWZ coset construction. To see this, let us work out the action near group identity $e$, where we will decompose Maurer-Cartan form $\theta_g= g^{-1}dg$ locally, see Eq.\,(\ref{mc1form}), along vertical space $\mf h$ and a horizontal space complimentary to $\mf h$. 

Firstly, for a connected, compact and semi-simple Lie group $G$ with its closed subgroup $H$, the coset space $M$ is $reductive$ homogeneous space, i.e. the Lie algebra $\mathfrak{g}$ of $G$ can be decomposed as
\beq
\mathfrak{g}=\mathfrak{h}\oplus\mathfrak{m}\,,
\label{decomposition}
\eeq
where $\mathfrak{h}$ is the subalgebra corresponding to the subgroup $H$, and $\mathfrak{m}$ is a transverse subspace that is preserved by the adjoint action of $H$, i.e.,
\beq
{\rm ad}_H\mathfrak{m}=\mathfrak{m}\,.
\label{AdjH}
\eeq
In principle, subspace $\mf m$ complimentary to $\mf h$ is quite arbitrary. However, similar to the discussion of ${\rm CP}^{N-1}$ embedded into $S^{2N-1}$\cite{CCSV2}, we can utilize the Killing form $K$, see Eq.\,(\ref{Killing}), to define
$$\mf m=\mf h^\perp\ ,$$
so that homogeneous space $M$ is a Riemann submersion of $G$, and the tangent space $\T_o M$, with $o\equiv\pi(e)$, is identified with $\mf m$. Under this decomposition, for $H_i\in\mathfrak{h}$ and $X_a\in\mathfrak{m}$, we have
\beqn
\Tr(H_iH_j)=-\delta_{ij},\ \Tr(X_aX_b)=-\delta_{ab},\ {\rm and}\ \Tr(H_iX_a)=0\,.
\label{perp decomp}
\eeqn
Now $\theta_g$ is decomposed as
\beqn
\theta_g(\phi)=e_g(\phi)+\omega_g(\phi)\equiv e_g^a X_a+\omega_g^i H_i\,,
\label{decomp mc1form}
\eeqn
where $e_g$ are called \emph{basic forms} and $\omega_g$ is \emph{canonical connection} for bundle $\pi: G\rightarrow M$. 

Now we can use gauge fields to eliminate redundant degrees of freedoms. For CCWZ construction, the unitary  gauge is chosen to remove all Nambu-Goldstone bosons on $\mf h$, i.e.,
\begin{equation*}
g(\phi)={\rm Exp}(\delta^a_\alpha\phi^{\alpha}X_a),
\end{equation*}
on a local chart $\{\phi^\alpha\in U_s\subset M\}$ near $o\in M$. Such a choice, geometrically speaking, is equivalent that we specify a local section $s: U_s\subset M\rightarrow G$\,,
\beq
s(\phi)\equiv g(\phi)={\rm Exp}(\delta^a_\alpha\phi^{\alpha}X_a)\,.
\label{unitary gauge}
\eeq
Therefore, one can use $s^\ast\!:\!\T^\ast G\rightarrow\T^\ast M$ pullback basic forms $e_g$ to $M$\,,
$$e_\phi\equiv s^\ast e_g=e^a_{\ \alpha}d\phi^\alpha X_a\,,$$
and thus define the vielbein one-form $e_\phi$ on $M$. Similarly, canonical connection $\omega_g$ is also pullback: 
$$\omega_\phi\equiv s^\ast\omega_g=\omega^i_{\alpha}d\phi^\alpha H_i$$
as connection one-form locally defined on $M$.

After fixing the gauge by Eq.\,(\ref{unitary gauge}), we localized the Lagrangian on a local chart $\{\phi^\alpha\in U_s\subset M\}$:
$$S_{M}=-\frac{1}{2\lambda^2}\int_\Sigma\dx\,{\rm Tr}(e^a_\mu X_a+\omega^i_\mu H_i-A^i_\mu H_i)^2\, ,$$
where 
$$e^a_\mu=e^a_{\ \alpha}\partial_\mu\phi^\alpha,\ \ \omega^i_\mu=\omega^i_{\ \alpha}\partial_\mu\phi^\alpha$$
are vielbeins and connection one-form further pullback to spacetime $\Sigma$ by the map $\phi^\alpha: \Sigma\rightarrow U_s\subset M$.

Since gauge fields $A_\mu$ classically is non-dynamical, one can solve and express them in terms of goldstone fields $\phi^\alpha$ by equations of motion, and we get
\beq
A^i_\mu=\omega^i_\alpha\partial_\mu\phi^\alpha\,.
\label{Hgauge}
\eeq
Getting this expression back to action Eq.\,(\ref{L_G/H}), we find the action $S_{M}$ by CCWZ construction,
\beq
\begin{split}
S_{M}&=-\frac{1}{2\lambda^2}\int_\Sigma\dx\,{\rm Tr}(e^a_{\ \alpha}X_a e^b_{\ \beta} X_b)\partial_\mu\phi^\alpha\partial^{\mu}\phi^\beta\\[2mm]
&=\frac{1}{2\lambda^2}\int_\Sigma\dx\,\delta_{ab}e^a_{\ \alpha}e^b_{\ \beta}\partial_\mu\phi^\alpha\partial^{\mu}\phi^\beta\equiv\frac{1}{2\lambda^2}\int_\Sigma\dx\,\gamma_{\alpha\beta}\partial_\mu\phi^\alpha\partial^{\mu}\phi^\beta\,.
\end{split}
\label{L_G/H local}
\eeq
With that
$$\gamma_{\alpha\beta}=\delta_{ab}e^a_{\ \alpha}e^b_{\ \beta},\ {\rm for}\ a, b, \alpha , \beta =1,2...{\rm dim}\ \mathfrak{m}$$
is the metric on $M$ and $e^a_{\ \alpha}$ is its vielbeins correspondingly.\\

\subsection{\boldmath{$\mathcal N=(0, 1)$} supersymmetric sigma model on \boldmath{$M$}}

In this subsection we will supersymmetrize the action of sigma model on $M=G/H$, see Eq.\,(\ref{L_G/H}). In two-dimensional spacetime, we have Weyl-Majorana Grassmannian variable $\theta_R$ which helps form the smallest representation of supersymmetry, i.e. (0, 1) supersymmetry. The superderivative in superspace is defined as
$$D_L=-i\,\frac{\partial}{\partial\theta_R}-\theta_R\partial_{LL}$$
satisfying
$$\{D_L, D_L\}=2D_L^2=2i\partial_{LL}$$
where $\partial_{LL}$ denotes the partial derivative along light-cone coordinate $x_L$, and $\partial_{RR}$ for that of $x_R$ in what follows. The integration over Grassmannian variable $\theta_R$ is equal to differentiation:
$$\int{\rm d}\theta_R=\frac{\partial}{\partial\theta_R}=iD_L\vert_{\theta_R=0}\ .$$

An ordinary bosonic field $\phi$ will be promoted to its superversion $\Phi$, which is consisted of $\phi$ and a left-moving fermion $\psi_L$:
$$\Phi=\phi+i\theta_R\psi_L$$
To supersymmetrize the action Eq.\,(\ref{L_G/H}), beside scalar superfield $g(\Phi)$, we also need $(0, 1)$ supergauge multiplets $\{\mathcal V_L,\ \mathcal V_{RR}\}$ \cite{Gates:1987}. It is true that one can directly supersymmetrize the local form of Lagrangian in Eq.\,(\ref{L_G/H local}), which is already localized on certain patch of $M$, without introducing any auxiliary gauge fields. However, with the help of gauge fields, it is quite easy to track the information of isometric transformations on different local charts, and also facilitate discussion of holonomy anomalies in next section.

The (0, 1) supergauge potential $\{\mathcal V_L,\ \mathcal V_{RR}\}$ are given as
\beq
\begin{split}
&\mathcal V_L=\eta_L-\theta_R A_{LL}\,,\\[1mm]
&\mathcal V_{RR}=A_{RR}+i\theta_R\chi_R\,.
\end{split}
\eeq
Under supergauge transformation
\beq
\begin{split}
&\mathcal V_L\rightarrow \mathcal H^{-1}\mathcal V_L\mathcal H+\mathcal H^{-1}D_L\mathcal H\,,\\[1mm]
&\mathcal V_{RR}\rightarrow \mathcal H^{-1}\mathcal V_{RR}\mathcal H+\mathcal H^{-1}\partial_{RR}\mathcal H\,,
\end{split}
\label{supergauge}
\eeq
where $\mathcal H$ is an arbitrary scalar superfield, one can remove field $\eta_L$ by choosing Wess-Zumino gauge. After this choice of supergauge  the residual is normal gauge transformations on gauge field $A_\mu=(A_{LL},A_{RR})$ and gaugino field $\chi_R$,
\beq
\begin{split}
&A_{\mu}\rightarrow h^{-1}A_{\mu}h+h^{-1}\partial_{\mu}h\,,\\[1mm]
&\chi_R\rightarrow h^{-1}\chi_Rh\,,
\end{split}
\eeq
where field $h$ is the bosonic component of superfield $\mathcal H$.

Now we have all ingredients needed to supersymmetrize Lagrangian Eq.\,(\ref{L_G/H}). We promote bosonic field $g(x)$ to be scalar superfield $\mathcal G(x, \theta_R)$ taking values on group $G$. The bosonic part of $\mathcal G$ is $g(x)$ while fermionic part is defined such as
\beq
\psi_L=\psi^A_LF_A\equiv \mathcal G^{-1}D_L\mathcal G\vert_{\theta_R=0}\,,
\label{fermion def}
\eeq
and thus,
$$\mathcal G=g+i\theta_R g\psi_L^AF_A\,,$$
where $F_A$ are the generators of Lie algebra $\mf g$ in fundamental representation as before. Under this definition, the fermionic action of $S^{(0, 1)}_{M}$ becomes canonical. Gauge fields $A_\mu$ are also enhanced to $\{\mathcal V_L, \mathcal V_{RR}\}$ taking values on Lie algebra $\mf h$.\\

The $(0, 1)$ supersymmetric action now written in superspace is given as
\beqn
S^{(0, 1)}_{M}=\frac{i}{2}\int_\Sigma\dx\int{\rm d}\theta_R{\rm Tr}\big[(\mathcal G^{-1}D_L\mathcal G-\mathcal V_L)(\mathcal G^{-1}\partial_{RR}\mathcal G - \mathcal V_{RR})\big].
\label{SS1}
\eeqn
Superfield $\mathcal G$ admits a $\mathcal H$ super-gauge transformation as designed,
$$\mathcal G\rightarrow \mathcal G\mathcal H\,.$$
To obtain the action in components, we impose Wess-Zumino gauge to remove $\eta_L$,
$$\mathcal V_L=-\theta_RA_{LL}\,.$$
Integrating $\theta_R$ out, we get
\beq
\begin{split}
S^{(0,1)}_{M}=& -\frac{1}{2}\int_\Sigma\dx\,\Tr[(g^{-1}\partial_{LL}g-A_{LL})(g^{-1}\partial_{RR}g-A_{RR})]\\[1mm]
&-\frac{i}{2}\int_\Sigma\dx\,\Tr[\psi_L(\partial_{RR}+g^{-1}\partial_{RR}g+A_{RR})\psi_L]\\[1mm]
&-\frac{i}{2}\int_\Sigma\dx\, \Tr(\chi_R\psi_L)\,.
\end{split}
\label{superL_G/H}
\eeq
The action still has ordinary $H$-gauge invariance,
\beq
\begin{split}
&g\rightarrow gh\,,\ \ \psi_L\rightarrow h^{-1}\psi_L h\,;\\[1mm]
&A_\mu\rightarrow h^{-1}A_\mu h+h^{-1}\partial_\mu h\,,\\[1mm]
&\chi_R\rightarrow h^{-1}\chi_R h\,.
\end{split}
\label{H-gauge trans}
\eeq
As before we decompose $g^{-1}\partial_\mu g$ and $\psi_L$ along horizontal and vertical directions,
\beq
\begin{split}
&g^{-1}\partial_\mu g=e^a_\mu X_a+\omega^i_\mu H_i\,,\\[1mm]
&\psi_L=\mathcal G^{-1}D_L\mathcal G\vert_{\theta_R=0}=\psi_L^aX_a+\psi_L^iH_i\,.
\end{split}
\label{perp decomp 2}
\eeq
Since $A_\mu$ and $\chi_R$ are non-dynamical, we solve these constraints by varying $A_\mu$ and $\chi_{R}\,$, and have
\beq
\begin{split}
&A^i_{RR}=\omega^i_{RR}\,,\\[1mm]
&A^i_{LL}=\omega^i_{LL}+\frac{i}{2}\,C^i_{ab}\psi_L^a\psi_L^b\,,\\[1mm]
&\psi_L^i=0\,,
\end{split}
\label{eom Apsi}
\eeq
where we have used Eq.\,(\ref{perp decomp}), the anti-symmetric property of $\psi_L^a$, and the commutator relations,
\beqn
[H_i, H_j]=C^k_{ij}H_k\,,\ \ [H_i, X_a]=C^c_{ia}X_c\,,\  \  [X_a, X_b]=C^k_{ab}H_k+C^c_{ab}X_c\ .
\label{commutator2}
\eeqn
From the first two formulas in Eq.\,(\ref{commutator2}) above, we see that, under this decomposition, Lie subalgebra $\mf h$ \emph{reducibly} acts on $\mf g$, or say, the adjoint representation of $\mf g$ restricted to $\mf h$ is decomposed as
\beq
({\rm ad}\ {\mf g})\vert_{\mf h}={\rm ad}\ {\mf h}\oplus\varrho\,,
\label{restrict}
\eeq
where $\varrho$ denotes the representation of $\mf h$ acting on subspace $\mf m$. We will see soon that this observation is very important to determine if anomalies produced by chiral fermions can be removed, and for us to write the most general action.  

Substituting  Eq.\,(\ref{eom Apsi}) back to action (\ref{superL_G/H}), we have
\beq
\begin{split}
S^{(0,1)}_{M}&=~~\frac{1}{2}\int_\Sigma\dx\ \delta_{ab}e^a_{LL}e^b_{RR}\\[1mm]
&~~+\frac{i}{2}\int_\Sigma\dx\ \psi^a_L(\partial_{RR}\delta_{ac}+\omega_{RR}^iC_{aic}+\frac{1}{2}e^b_{RR}C_{abc})\psi^c_L\ .
\end{split}
\eeq
It is not the final result yet because we should assign coupling constants $\lambda^2$, which is related to how vielbein $e^a_{\mu}$ and fermion $\psi_L^a$ transforms under gauge transformation. From Eq.\,(\ref{H-gauge trans}) and (\ref{perp decomp 2}), writing the transformations in components:
\beq
\begin{split}
&e^a_\mu\rightarrow\rho(h^{-1})^a_{\ b}e^b_\mu\,,\ \ \ \psi_L^a\rightarrow\rho(h^{-1})^a_{\ b}\psi_L^b\,;\\[1mm]
&\omega^i_\mu C^a_{ib}\equiv\omega^a_{\mu b}\rightarrow\left(\rho(h^{-1})\omega_\mu\rho(h)+\rho(h^{-1})\partial_\mu\rho(h)\right)^a_{\ b}\,,
\end{split}
\label{H-gauge trans rho}
\eeq
where $\rho$ denotes the $H$-isotropy representation on $\mf m$ corresponding to $\varrho$, i.e.,\footnote{\,Since we chose normalized and orthogonal bases $\{X_a\}$, $\rho$ is in fact orthogonal real representation of $H$ on $\mf m$, i.e. $\rho(h)^a_{\ b}=\rho(h^{-1})_b^{\ a}$, by which Eq.\,(\ref{H-gauge trans rho}) can be verified.}
\beq 
h^{-1}X_ah\equiv\rho(h)_a^{\ b}X_b\  \ {\rm for}\ X_{a,b}\in\mf m\,.
\eeq
Equation \,(\ref{H-gauge trans rho}) implies that the tangent bundle $\T M$ is identified to the associated $H$-principal fiber bundle with vector space $\mf m\,$,
\beq
\T M\simeq G\times_{\varrho}\mf m\ ,
\label{AVH}
\eeq
on which vielbeins $e^a$ and fermions $\psi^a_L$ are the basic form, and $\omega^a_{\ b}$ is the connection in $\varrho$ representation. Now if $\rho$, or equivalently $\varrho$\,, is further reducible on $\mf m$\,,
$$\rho=\bigoplus_{r_a}\rho_{r_a}\ ,$$
we can assign different coupling constant $\lambda_a^2$ to each independent representation\,\footnote{\,If there exists right isometries after we gauge out $H\subset G_R$, the number of coupling constants will be as many as the independent representation of normalizer of $H$. For more details, we refer readers to reference\cite{Castellani:1984}.} $r_a$ of $H$ on $\mf m$.
Based on the argument above, we rescale vielbein $e^a_{\mu}$ and fermion $\psi_L^a$ in respect to the representations they belong to,
\beq
e^a_{\mu}\rightarrow\frac{1}{\lambda_a}\,e^a_\mu,\ \ \psi^a_{L}\rightarrow\frac{1}{\lambda_a}\,\psi^a_L\,,
\label{scales}
\eeq
and the action changes to 
\beq
\begin{split}
S^{(0,1)}_{M}&=\frac{1}{2\lambda^2_a}\int_\Sigma\dx\ \delta_{ab}e^a_{LL}e^b_{RR}\\[1mm]
&+\frac{i}{2\lambda^2_a}\int_\Sigma\dx\ \psi^a_L\Big(\partial_{RR}\delta_{ac}+\omega_{RR}^iC_{aic}+\frac{\lambda_a}{2\lambda_b\lambda_c}\,e^b_{RR}C_{abc}\Big)\psi^c_L\,,
\end{split}
\eeq
where we used the fact that connection $\omega^i_{RR}C_{aic}$ is block diagonal and thus indexes $a$ and $c$ are forced in the same representation, say $\lambda_a=\lambda_c$. Further, anticommutativity of fermions $\psi^{a, c}_L$ requires us to antisymmetrize the indexes  $a$ and $c$ of term\,\footnote{\,$\tau$ and $\kappa$ are respectively the torsion and contorsion of homogeneous spaces $M$, see also in \cite{Castellani:1984}.} $\frac{\lambda_a}{2\lambda_b\lambda_c}\,C_{abc}$\,,
$$\tau_{abc}\equiv-\frac{1}{2}\frac{\lambda_a}{\lambda_b\lambda_c}C_{abc}\rightarrow\kappa_{abc}\equiv\frac{1}{2}\Big(\frac{\lambda_b}{\lambda_a\lambda_c}-\frac{\lambda_c}{\lambda_b\lambda_a}-\frac{\lambda_a}{\lambda_b\lambda_c}\Big)C_{abc}\,.$$
We finally have
\beq
\begin{split}
S^{(0,1)}_{M}&=\frac{1}{2\lambda^2_a}\int_\Sigma\dx\ \delta_{ab}e^a_{LL}e^b_{RR}\\[1mm]
&~~~+\frac{i}{2\lambda^2_a}\int_\Sigma\dx\ \psi^a_L\big(\partial_{RR}\delta_{ac}+\omega_{RR}^iC_{aic}-e^b_{RR}\kappa_{abc}\big)\psi^c_L\\[1mm]
&=\frac{1}{2\lambda^2_a}\int_\Sigma\dx \Big[ \delta_{ab}e^a_{LL}e^b_{RR}+i\psi^a_L\big(\partial_{RR}\delta_{ac}+\widetilde\omega_{RRac}\big)\psi^c_L\Big]\,,
\end{split}
\label{superL_G/H 2}
\eeq
where $\kappa$ term is absorbed into connection $\omega$ to define:
$$\widetilde\omega_{ac}\equiv\omega_{ac}-\kappa_{ac}\ ,$$
as the Levi-Civita connection of homogeneous spaces $M$. Since field $\kappa_{ab}$ is tensorial, under $H$-gauge transformation, we still have
\beq
\begin{split}
&e^a_\mu\rightarrow\rho(h^{-1})^a_{\ b}e^b_\mu\,,\ \ \ \psi_L^a\rightarrow\rho(h^{-1})^a_{\ b}\psi_L^b\,,\\[1mm]
&\widetilde\omega^a_{\mu b}\rightarrow\left(\rho(h^{-1})\tilde\omega_\mu\rho(h)+\rho(h^{-1})\partial_\mu\rho(h)\right)^a_{\ b}\,.
\end{split}
\eeq\\

\subsection{Isometry anomalies of sigma model on \boldmath{$M$}}
\label{2.3}
In this subsection, we will disclose the relation between isometric and $H$-gauge transformations, see Eq.\,(\ref{H-gauge trans rho}), and then calculate isometry anomalies of the action Eq.\,($\ref{superL_G/H 2})$. For brevity, in what follows, including also the next section, we will only label one, instead of two, $``R\,"$ or $``L"$ as the subscription of all quantities when it leads to no confusion. \\

\noindent Now let us consider isometries of the action. We start from the fiberation:
$$\xymatrix{H \ar[r]^i & G \ar[r]^\pi & M}\ ,$$
that all (left) isometries $l_k: M\rightarrow M$ are induced from left translations\,\footnote{\,As mentioned, there may be also right isometries on $M$ induced by right translation on $G$ if the normalizer of $H$ is larger than $H$ itself. There are also corresponding right isometry anomalies, but the discussion of them are similar to that of the left. More detailed can be found in \cite{AlvarezGaume:1985yb}.} $L_k$:
\beq
L_k: g(x)\mapsto kg(x),\ \ {\rm for}\ k\in G\,,
\label{Lk}
\eeq
and we have the following commuting diagram:
\beq
\xymatrix{&G \ \ar_{\pi}[d] \ar^{L_k}[r] &G \ \ar^{\pi}[d]\nonumber
\\&M \ \ar^{l_k}[r] &M}\nonumber
\eeq
\beq \hspace{1cm} {\rm with}\ \pi\circ L_k=l_k\circ\pi \label{G/Hisotransf}\eeq
It is easily seen that these left translations keep action Eq.\,(\ref{superL_G/H}) invariant trivially since $k\in G$ is a constant group element. \\

When investigating isometric transformation $l_k$ on $M$, we are required to choose a local trivialization, or say, a local section $s: U_s\subset M\rightarrow G$. Physically speaking, we fix a gauge, for example the CCWZ coset construction where unitary gauge is chosen (see Eq.\,(\ref{unitary gauge})), and localize the action $S^{(0, 1)}_M$ on $U_s$ by the coordinates $\{\phi^\alpha\}\in U_s\subset M$. More explicitly, we have
\beq
g=s(\phi)\ .
\label{s section}
\eeq
Therefore, vielbeins $e^a_\mu$, connection $\omega^a_{\mu b}$ as well as fermions $\psi^a_L$ are pullback to $U_s\subset M$ and expressed by Eqs.\,(\ref{s section}) and (\ref{perp decomp 2}) as
\beqn
&&s^\ast(e^a_\mu)=e^a_{\ \alpha}\partial_\mu\phi^\alpha=-\Tr\Big(X^as^{-1}\frac{\partial s}{\partial\phi^\alpha}\Big)\partial_\mu\phi^\alpha\,,\nonumber\\
&&s^\ast(\omega^a_{\mu b})=\omega^a_{\alpha b}\partial_{\mu}\phi^\alpha=-\Tr\Big(H^is^{-1}\frac{\partial s}{\partial\phi^\alpha}\Big)\partial_\mu\phi^\alpha C^a_{ib}\,,\\
&&s^\ast(\psi_L^a)=-\Tr\Big(X^as^{-1}\frac{\partial s}{\partial\Phi^\alpha}\Big)D_L\Phi^\alpha\vert_{\theta_R=0}=e^a_{\ \alpha}\psi^\alpha_L \,.\nonumber
\label{pullback}
\eeqn
From now on, we will not label $s^\ast$ to distinguish these quantities as forms on bundle $G\times_\rho\mf m$ or locally pullback to $U_s\subset M$. It should lead no confusion in contexts. Thanks to the gauge fixing, the action localized on $U_s$ is given as
\beq
\begin{split}
S^{(0,1)}_{U_s}[\phi, \psi_L]&=\frac{1}{2\lambda^2_a}\int_\Sigma\dx\ \delta_{ab}e^a_{\ \alpha}e^b_{\ \beta}\partial_{L}\phi^\alpha\partial_{R}\phi^\beta\\[1mm]
&+\frac{i}{2\lambda^2_a}\int_\Sigma\dx\ \psi^a_L\big(\partial_{R}\delta_{ac}+\partial_{R}\phi^\alpha\tilde\omega_{\alpha ac}\big)\psi^c_L\,.
\end{split}
\label{local superL_G/H}
\eeq
This action should be invariant under isometric transformation
\beq
l_k: \phi\mapsto l_k(\phi)\,.
\eeq
We will show that vielbeins, connections and fermions are transformed under $l_k$ as a special type of $H$-gauge transformation, see Eq.\,(\ref{H-gauge trans rho}). Then the invariance of action (\ref{local superL_G/H}) is guaranteed.

To see this, one can directly calculate their Lie derivatives respect to isometries $l_k$ (cf.\cite{AlvarezGaume:1985yb} for example). Here instead, we interpret this issue in language of fiber bundle, which we presented and explained in great details for ${\rm CP}^{N-1}$ case in \cite{CCSV2}. For a given section $s$, or a fixed gauge, we map the local patch $U_s$ to $G$ by
$$s(\phi)=g\in G\,.$$
A left translation $L_k$ acting on $s(\phi)$ not only induces isometric transformation $l_k$ on chart $\{\phi^\alpha\}$ but also changes the fixed gauge. When we consider the isometric transformations of quantities $e^a_\mu$, $\omega^a_{\mu b}$ and $\psi_L^a$ under the original fixed gauge, we are required to accompany them by a $H$-gauge transformation $h(\phi, k)$ to compensate the change:
\beq
L_ks(\phi)h(\phi, k)=s(l_k(\phi)),\ \ {\rm for}\ k\in G\,.
\label{ss'}
\eeq
Or equivalently to say, the composition of $L^{-1}_k\circ s\circ l_k$ define another section $s^\prime$, see the commuting diagram:
\beq
\xymatrix{& G & G\ar_{L_k^{-1}}[l]\nonumber
\\\ \ & U_{s^\prime} \ar_{s^\prime}[u] \ar^{l_k}[r] & U_s \ar_s[u]\nonumber}
\eeq
Sections $s^\prime$ and $s$ are related by a $H$-gauge transformation $h(\phi, k)$, i.e. Eq.\,(\ref{ss'}),
$$s^\prime(\phi)=s(\phi)h(\phi, k)\,.$$
Now, after isometric transformation $l_k$, vielbeins, connections and fermions are pullback to $U_{s^\prime}$ by $s^{\prime\ast}$ and are related to those pullback by $s^\ast$ as
\beq
\begin{split}
&e^a_\mu\rightarrow e^{\prime a}_\mu=\rho(h^{-1}_{\phi, k})^a_{\ b}e^b_\mu,\ \ \ \psi_L^a\rightarrow\psi_L^{\prime a}=\rho(h^{-1}_{\phi, k})^a_{\ b}\psi_L^b\,,\\[1mm]
&\omega^{a}_{\mu b}\rightarrow\omega^{\prime a}_{\mu b}=\left(\rho(h^{-1}_{\phi, k})\omega_\mu\rho(h_{\phi, k})+\rho(h^{-1}_{\phi, k})\partial_\mu\rho(h_{\phi, k})\right)^a_{\ b}\,,
\end{split}
\label{Hgauge sp}
\eeq
where $h_{\phi, k}\equiv h(\phi, k)$ for short. Infinitesimally one can expand,
$$l_k\simeq 1+\epsilon^A K_A(\phi)\,, \ L_{k^{-1}}\simeq 1-\epsilon^A F_A\,, \ {\rm and}\ \ h(\phi, k)\simeq 1+\alpha^i(\phi, \epsilon)H_i\,,$$
and get them back to Eq.\,(\ref{ss'}) to explicitly solve $K_A$\,, the Killing field for isometries $l_k$\,, and $\alpha^i$. However it is unnecessary to know their explicit expression. We only need to know, infinitesimally, 
\beq
\begin{split}
&\delta_\alpha e^{a}_\mu=-\varrho(\alpha)^a_{\ b}e^b_\mu\,,\  \ \delta_\alpha\psi_L^{a}=-\varrho(\alpha)^a_{\ b}\psi_L^b\,,\\[1mm]
&\delta_\alpha\omega^{a}_{\mu b}=\partial_\mu\varrho(\alpha)^a_{\ b}+[\omega_\mu,\ \varrho(\alpha)]^a_{\ b}\,,
\end{split}
\label{iso}
\eeq
where 
$$\varrho(\alpha)^a_{\ b}\equiv\alpha^i\varrho(H_i)^a_{\ b}=\alpha^iC^a_{ib}\,.$$
One can further show that contortion $\kappa^a_{\ b}$ transforms tensorially,
$$\delta_\alpha\kappa^{a}_{\mu b}=[\kappa_\mu,\ \varrho(\alpha)]^a_{\ b}\ \ {\rm and\ thus}\ \ \delta_\alpha\widetilde\omega^{a}_{\mu b}=\partial_\mu\varrho(\alpha)^a_{\ b}+[\widetilde\omega_\mu,\ \varrho(\alpha)]^a_{\ b}\ .$$\\
\noindent Now, for isometry anomalies, we use action $S^{(0, 1)}_{U_s}[\phi, \psi_L]$ to calculate the effective action. Similarly to the discussion in \cite{CCSV2}, anomalies are only produced from fermionic integration effective action. We thereby integrate out the fermionic part of action Eq.\,(\ref{local superL_G/H}) and have
\beq
i\,\mathcal W^{s}_{f}[\widetilde\omega_{R}]=\frac{i}{16\pi}\int_\Sigma\dx\ \Tr(\widetilde\omega_{R}\,\frac{\partial_{L}\partial_{L}}{\partial^2}\,\widetilde\omega_{R})+\mathcal O(\widetilde\omega^3_{R})\,,
\eeq
where the superscript $s$ denotes that our perturbative calculation is performed on the local chart $U_s$. Varying $\mathcal W^{s}_{f}$\,, we produce isometry anomalies $\mathcal I_{\alpha}$\,,
\beq
\mathcal I_{\alpha}=\delta_{\alpha}\mathcal W^{s}_{f}=-\frac{1}{8\pi}\int_\Sigma\dx\ \Tr(\alpha \partial_{L}\widetilde\omega_{R})\,.
\label{anomalyG/H}
\eeq

To conclude, in this chapter we have calculated isometry anomalies of generic (0,1) supersymmetric sigma models defined on manifold $M=G/H$. To perform perturbative calculation, we need to specify a local chart $U_s$ on $M$ to define the model and thus a section $s$ from $U_s$ to $G$. After integrating out fermions, we find the effective action $\mathcal W_{f}$ which is also defined on the local patch $U_s$. However in many cases the effective action does not bear isometries $l_k$\,, for $k\in G$, it had before and thus produces isometry anomalies. We established a correspondence between the isometry anomalies and some specific $H$-gauge anomalies when considering to define the effective action $\mathcal W_{f}$ on the intersection of two local patch $U_s\cap U_{s^\prime}$, where local patch $U_{s^\prime}=l^{-1}_k(U_s)$ induced by isometries. This observation actually inevitably leads us to consider anomalies not only localized on a specified coordinates or local chart, but also to evaluate if the effective action $\mathcal W_{f}$ can be consistently defined on different local patches and their intersections. If so, one is able to transit $\cw_{f}$ from patch to patch without producing any $H$-gauge anomalies, or called holonomy anomalies. Since isometry anomalies is a specific type of $H$-gauge anomalies, they will for sure vanish in this situation. Otherwise when a model is suffered from holonomy anomalies, it is not even possible to globally define the theory quantum mechanically, and thus makes no sense to consider its isometry anomalies. Therefore in the next chapter, we will focus on holonomy anomalies and the criteria when they vanish or can be canceled by counterterms.\\
\newpage

\section{Holonomy anomalies}
\label{HA}
In last section, we have argued that, to define sigma models on Homogeneous spaces $M=G/H$, it is required that, prior to consider isometry anomalies, the theory should be independent of the choices of sections $s: U_s\subset M\rightarrow G$, or say the ease of holonomy anomalies. The holonomy anomalies will arise when we change from a section $s$ to another $s^\prime$, or physically speaking, from a fixed gauge to another. Therefore they correspond to an arbitrary $H$-gauge transformation, see Eq.\,(\ref{H-gauge trans}) and Eq.\,(\ref{H-gauge trans rho}), while isometry transformations are a special type of $H$-gauge transformation, see Eq.\,(\ref{Hgauge sp}). Therefore, once holonomy anomalies are removed, isometry anomalies will automatically vanish as well. We will thus focus ourselves on holonomy anomalies and their cancellation condition.

\subsection{Anomaly matching condition}
\label{3.1}
From Eq.\,(\ref{anomalyG/H}), we know that $\alpha$ and $\widetilde\omega_R$ are taking values in the $\varrho$ representation of $\mf h$ Lie subalgebra. On the other hand, we will show that counterterms that can be introduced is in the $F(\mf g\vert_{\mf h})$ representation, say the fundamental representation $F$ of $\mf g$ representation restricted on $\mf h$. Choosing the fundamental representation $F$ is merely of convention, since we are free to redefine our coupling constants corresponding to other different representation. Roughly speaking, the counterterm we can introduce is an analog of gauged WZW term $\cw_{\rm c.t}[g, A]$, which is well-known to produce gauge anomalies when the gauge fields $A$ taking values in $\mf h$ are not in a ``safe" representation\cite{Witten:1992}. When certain matching condition on $\varrho$ and $F(\mf g\vert_{\mf h})$ is satisfied, these two anomalies are canceled. From now on to distinguish the difference between representation $\varrho$ and $F(\mf g\vert_{\mf h})$, we will use $\Tr_\varrho$ and $\Tr_F$ to label under which representation we take the trace.\\ 

First we will explore more on the structure of effective fermionic action $\cw_{f}$. In what follows, we will not fix ourselves in any specific gauge, and will not solve gauge field $A$ in terms of $g$ as Eq.\,(\ref{eom Apsi}), because it will help better track the information of gauge transformations on $\cw_f$ and, more importantly, give us an explicit expression of fermionic effective action. We thus use Eq.\,(\ref{superL_G/H}) and rewrite the fermionic part as
\beqn
S_f=-\frac{i}{2}\int_\Sigma\dx\ \Big[\Tr_F\ \psi_L(\partial_R\psi_L+[A_R, \psi_L]) +\Tr_F\ \psi_L(g^{-1}\partial_R g-A_{R})\psi_L\Big]
\label{S_f}
\eeqn
with
$$\psi_L=\psi_L^aX_a\,,\ \ \ {\rm and}\ \ A_R=A^i_RH_i\,.$$
The two parts of above equation are separately gauge invariant classically. However the second term, $g^{-1}\partial_R g-A_{R}$ coupling to fermions, transforms tensorially under a $H$-gauge transformation,
$$g^{-1}\partial_R g-A_{R}\rightarrow h^{-1}(g^{-1}\partial_R g-A_{R})h\ ,$$
while in the first term chiral fermions $\psi_L$ couple to gauge fields $A_R$ and will produce genuine anomalies. If we can find counterterms to offset the anomalies from the first term in Eq.\,(\ref{S_f}), the anomalies from the second one can be removed also by an analog of Bardeen like counterterm in two dimensions. Let us see how it works.

In fact we can ask more for an explicit structure on the anomalous part of $\cw_f$ in two-dimensional spacetime due to Polyakov and Wiegmann\cite{Polyakov:1983}. In two dimensions, one can parameterize gauge fields as
\beq
A_R=\tilde h^{-1}\partial_R\tilde h\ \ {\rm and}\ \ A_L=\tilde h^{\prime -1}\partial_L\tilde h^\prime\,,
\label{PW}
\eeq
where fields $\tilde h(x)$ and $\tilde h^\prime(x)$ are elements in $H$ and under gauge transformation
$$\tilde h\rightarrow\tilde h h\,,\ {\rm and}\ \ \tilde h^\prime\rightarrow \tilde h^\prime h\ .$$ 
Notice that, since $\tilde h\neq\tilde h^\prime$, $A_\mu$ are not flat connection. One can solve $\tilde h$ and $\tilde h^\prime$ in terms of the Wilson lines of $A_R$ and $A_L$, although the expressions is surly non-local,
$$\tilde h(x)=-P\,{\rm e}^{-{\int_{C_x}}{{\rm d}\xi_LA_R}},\ \ {\rm and}\ \ \tilde h^\prime(x)=-P\,{\rm e}^{-{\int_{C_x}}{{\rm d}\xi_RA_L}}\,,$$
where $C_x$ is a path from certain fixed point to $x$ and $P$ denote path ordered integral. With the help of $\tilde h$, one can explicitly write down the anomalous part of $\cw_f$. Let us first rewrite term $g^{-1}\partial_R g-A_{R}$ as
\beq
g^{-1}\partial_R g-A_{R}=g^{-1}\partial_R g-\tilde h^{-1}\partial_R \tilde h=g^{-1}\partial_R(g\tilde h^{-1})(g\tilde h^{-1})^{-1}g\ .
\label{varphi}
\eeq
Clearly $g\tilde h^{-1}$ is gauge invariant. Actually, if we redefine fermions $\psi_L$ as
\beq
\psi_L=\tilde h^{-1}\zeta_L\tilde h\ \ {\rm or\ in\ components}\ \ \psi^a_L=\rho(\tilde h^{-1})^a_{\ b}\zeta^b_L\ ,
\label{psi-zeta}
\eeq
the action $S_f$ changes to 
\beqn
S^\prime_f=-\frac{i}{2}\int_\Sigma\dx\ \Tr_F\Big[\zeta_L\partial_R\zeta_L+\zeta_L(g\tilde h^{-1})^{-1}\partial_R(g\tilde h^{-1})\zeta_L\Big].
\label{Sf'}
\eeqn
Notice now that both $\zeta_L$ and $g\tilde h^{-1}$ are gauge invariant. After integrating out $\zeta_L$, the effective fermionic action is guaranteed to be gauge invariant as well,
\beq
\cw^\prime_f[g\tilde h^{-1}]=-i{\rm log}\int\mathcal D\zeta_L\ {\rm e}^{iS^\prime_f}\,.
\label{W'}
\eeq
Therefore, we can interpret that the anomaly Eq.\,(\ref{anomalyG/H}) is raised in a functional determinant when we change fermionic measure,
$$\int\mathcal D\psi_L=\int\mathcal Det^{-1}\Big[\frac{\delta\psi_L}{\delta\zeta_L}\Big]\mathcal D\zeta_L\,,$$
and we will calculate the determinant above. The method we will use is mainly based on \cite{Polyakov:1983}. 

The determinant is an integrated version of anomaly Eq.\,(\ref{anomalyG/H}). Now since we keep gauge fields explicitly, the anomaly equation becomes:
\beq
\begin{split}
\mathcal A_{\alpha}&=-\frac{1}{8\pi}\int_\Sigma\dx\ \Tr_\varrho\alpha\,\partial_{L}\Big(\frac{1}{2}A_R+\frac{1}{2}\omega_R-\kappa\Big)\\[1mm]
&=-\frac{1}{8\pi}\int_\Sigma\dx\ \Tr_\varrho\alpha\partial_{L}A_R-\frac{1}{16\pi}\int_\Sigma\dx\ \Tr_\varrho\alpha\partial_L(\omega_R-A_R)
\end{split}
\label{anomalyM}
\eeq
where, in the second equality, we use
$$\Tr_\varrho\alpha\kappa=\frac{1}{2}\Big(\frac{\lambda_b}{\lambda_a\lambda_c}-\frac{\lambda_c}{\lambda_b\lambda_a}-\frac{\lambda_a}{\lambda_b\lambda_c}\Big)\alpha^ie^bC^c_{ia}C^a_{bc}=0\ ,$$
because of $\Tr_{\varrho, F} H_iX_b=0$, see Eq.(\ref{perp decomp}).
The first term in anomaly Eq.\,(\ref{anomalyM}) corresponds to the first part of action Eq.\,(\ref{S_f}):
\beq
S^1_{f}= -\frac{i}{2\lambda^2_a}\int_\Sigma\dx\ \Tr_F\ \psi_L(\partial_R\psi_L+[A_R, \psi_L])=\frac{i}{2\lambda^2_a}\int_\Sigma\dx\ \psi_{La}(\partial_R\psi^a_L+A^i_{R}C^{a}_{ib}\psi^b_L)\ ,\nonumber
\eeq 
where we write the action in components and rescale fermions to make the coupling constants explicit. For $A_R$ parameterized as $\tilde h^{-1}\partial_R\tilde h$, we now aim to find an effective action $\cw^1_{f}[\tilde h]$ which corresponds to $S^1_f$ and satisfies
$$\delta_\alpha\cw^1_{f}[\tilde h]=\mathcal A^1_\alpha=-\frac{1}{8\pi}\int_\Sigma\dx\ \Tr_\varrho (\tilde h^{-1}\delta\tilde h)\partial_{L}(\tilde h^{-1}\partial_R\tilde h)\ ,$$
where we also put $\alpha=\tilde h^{-1}\delta\tilde h$\,. Due to Polyakov and Weigmann, the effective action can be solved as
\beq
\cw_f^1=\cw_{\rm PW}[\tilde h]\equiv\frac{1}{16\pi}\int_\Sigma\dx\ \Tr_\varrho(\tilde h^{-1}\partial_R\tilde h)(\tilde h^{-1}\partial_L\tilde h)-\frac{1}{24\pi}{\int_{\tilde h(B)}\!\!\!\!\!\Tr_\varrho(\tilde h^{-1}d\tilde h)^3}\,,
\eeq
where, in the second term, $\tilde h=\tilde h(x, t)$ has been extended\,\footnote{\,The extension of $\tilde h(x, t)$, and also that of $g(x, t)$ later, are always assumed to exist. For situations when $\pi_2(H)$ or $\pi_2(G)$ are non-trivial, we will present in our future work on global anomalies.} to bulk $B$ bounded by $\Sigma$. It is well-known that the second term is multi-valued and can be rewritten as a local form on spacetime $\Sigma$, and thus we still have a local theory defined on $\Sigma$ rather than the bulk $B$.

Beside this part, there is also the second term left in anomaly Eq.\,(\ref{anomalyM}),
$$\mathcal A^2_\alpha=-\frac{1}{16\pi}\int_\Sigma\dx\ \Tr_\varrho\alpha\partial_L(\omega_R-A_R)\,.$$
We have argued that $g^{-1}\partial_R g-A_R$, as well as $\omega_R-A_R$, transform tensorially and thus do not produce anomalies themselves, unless they are coupled to gauge fields as probes. Therefore we can easily verify that, a Bardeen-like counterterm, 
\beq
\cw_f^2[\tilde h, \omega_R-A_R]=\frac{1}{16\pi}\int_\Sigma\dx\ \Tr_\varrho (\omega_R-A_R)(\tilde h^{-1}\partial_L\tilde h)\,,
\eeq
satisfies
$$\delta_\alpha\cw_f^2=\mathcal A^2_\alpha\ ,$$
and, thus, is the second part of the anomalous effective action.

Overall we explicitly solve the anomalous part of effective action $\cw_f$, and the whole effective action $\cw_f$ is given as
\beq
\begin{split}
\cw_f&=\cw_f^1[\tilde h]+\cw_f^2[\tilde h, \omega_R-A_R]+\cw^\prime_f[g\tilde h^{-1}]\\[1mm]
&=-\frac{1}{24\pi}{\int_{\tilde h(B)}\!\!\!\!\!\Tr_\varrho(\tilde h^{-1}d\tilde h)^3}+\frac{1}{16\pi}\int_\Sigma\dx\ \Tr_\varrho \omega_R(\tilde h^{-1}\partial_L\tilde h)+\cw^\prime_f[g\tilde h^{-1}]\,,
\end{split}
\label{S_f2}
\eeq
~\\

\noindent Now, based on the anomalous effective action above, we are seeking conditions and counterterms $\cw_{\rm c.t.}[g, A_R]$. The key hint from Eq.\,(\ref{S_f2}) is that we need an analog of term $\Tr_\varrho(\tilde h^{-1}d\tilde h)^3$. It should first have same gauge transformation rule as $\tilde h^{-1}d\tilde h$\,,
$$\tilde h^{-1}d\tilde h\rightarrow h^{-1}(\tilde h^{-1}d\tilde h)h+h^{-1}dh,\ \ {\rm for}\ \ \tilde h\rightarrow \tilde hh\ .$$
and secondly be able to pullback to spacetime $\Sigma$ to define our theory in two dimensions.
However the only ingredient we have to satisfy the two conditions is
$$\cw_{c.t.}\sim\Tr_F(g^{-1}dg)^3\ .$$
An infinitesimal $H$-gauge transformation, c.f. Eq.\,(\ref{H-gauge trans}), is given as:
$$\delta_\alpha (g^{-1}dg)=d\alpha^iH_i+[g^{-1}dg, \alpha^iH_i],\ \ {\rm for}\ \ \delta_\alpha g=g\alpha^iH_i\,,$$
where we explicitly display $\alpha$ above taking values in $F(\mf g\vert_\mf h)$. Therefore we have:
$$\delta_\alpha\Tr_F(g^{-1}dg)^3\sim\Tr\ \alpha d(g^{-1}dg)\sim \alpha^id\omega^j\Tr H_iH_j\,.$$
Since we have already normalized generators $H_i$ in Eq.\,(\ref{perp decomp}), as
$$\Tr_F H_iH_i=-\delta_{ij},\ \ {\rm for\ any}\ \ H_{i, j}\in \mf h$$
the \emph{anomaly matching condition}, under our conventions, is
\beqn
\Tr_\varrho H_iH_j=c\,\Tr_FH_iH_j=-c\,\delta_{ij},\ \ {\rm for\ any}\ \ H_{i, j}\in \mf h\,,
\label{amc}
\eeqn
with some constant $c$. So long as the anomaly matching condition is satisfied, we can construct the 
counterterms $\cw_{\rm c.t.}$ as
\beq
\cw_{\rm c.t.}=\frac{c}{24\pi}{\int_{g(B)}\!\!\!\!\!\Tr_F(g^{-1}dg)^3}-\frac{c}{16\pi}\int_\Sigma\dx\ \Tr_F A_R(g^{-1}\partial_Lg)\,.
\label{ct}
\eeq
One can verify that, when Eq.\,(\ref{amc}) is met,
$$\delta_\alpha\cw_{\rm c.t.}+\mathcal{A_\alpha}=0\,.$$
At last, combining Eq.\,(\ref{ct}) and Eq.\,(\ref{S_f2}), we would expect the modified fermi\-onic action $\cw_{\rm eff}$ is gauge invariant,
\beq
\cw_{\rm eff}=\cw_f+\cw_{\rm c.t.}=\frac{c}{24\pi}{\int_{g\tilde h^{-1}(B)}\!\!\!\!\!\!\!\!\!\!\Tr_F\Big[(g\tilde h^{-1})^{-1}d(g\tilde h^{-1})\Big]^3}+\cw^\prime_f[g\tilde h^{-1}]\,.
\label{Weff}
\eeq\\

\subsection{Comments on counterterms}
\label{3.2}

So far we derived the anomaly matching condition Eq.\,(\ref{amc}), based on which the gauge invariant effective action, Eq.\,(\ref{Weff}), is constructed above. There are some interesting results and comments we want to put.\\

\noindent {\bf i. Anomaly matching condition}\\ 

The anomaly matching condition is a group theoretical result. In principle, if we understand how a subgroup $H$ is embedded to $G$, we can determine, by Eq.\,(\ref{amc}), whether a minimal $(0, 1)$ supersymmetric sigma model can be well-defined. Actually the statement is topological, when we will show in subsection $\ref{3.4}$, that Eq.\,(\ref{amc}) will be satisfied if and only if the \emph{first Pontryagin form} of $M$ vanishes, i.e. $p_1(M)=0$.\\ 

\noindent {\bf ii. \boldmath{$\cw_{\rm eff}$} incorporated with \boldmath{$(0, 1)$} supersymmetry}\\

Till now, besides requiring $(0, 1)$ supersymmetry on model building, we did not fully consider the role supersymmetry may play in the game. The counterterm $\cw_{\rm c.t.}$ we added is apparently non-supersymmetric, but it is required to define our theory. Now we want to proceed one step more, when we find the gauge invariant fermionic action $\cw_{\rm eff}$. For brevity, we use $\varphi\equiv g\tilde h^{-1}$ as the gauge invariant field, and $\cw_{\rm eff}$ is rewritten as
$$\cw_{\rm eff}[\varphi]=\frac{c}{24\pi}{\int_{\varphi(B)}\!\!\!\!\!\Tr_F(\varphi^{-1}d\varphi)^3}+\cw^\prime_f[\varphi]\,.$$
The second term is due to a path integral over fermions $\zeta_L$, see Eq.(\ref{Sf'}) and Eq.\,(\ref{W'}), 
$${\rm e}^{i\cw^\prime_f[\varphi]}=\int\mathcal D\zeta_L\exp\int_\Sigma\dx\ \Big(\!-\frac{i}{2\lambda^2}\Big)\Tr_F\left(\zeta_L\partial_R\zeta_L+\zeta_L\varphi^{-1}\partial_R\varphi\zeta_L\right)\,,$$
which has its supersymmetric counterpart $S_{b}$, see Eq.\,(\ref{L_G/H}). On the other hand, the first term, as a combination of anomalous and anomaly-counterterms, 
$$\cw_{\rm WZW}\equiv\frac{c}{24\pi}{\int_{\varphi(B)}\!\!\!\!\!\Tr_F(\varphi^{-1}d\varphi)^3}\,,$$
has no its supersymmetric pair. Therefore we will supersymmetrize this term. Actually the $\mathcal N=(1, 1)$ supersymmetrization of $\cw_{\rm WZW}$ is well-known in literatures back to 80's, c.f. \cite{Vecchia:1985} and \cite{Rohm:1984} for example. Here we do the similar to equip $\cw_{\rm WZW}$ with a $\mathcal N=(0, 1)$ supersymmetry.  Since field $\varphi$ is now gauge invariant, its $(0, 1)$ super-partner is also gauge invariant, and thus must be $\zeta_L$. The supersymmetrization of $\cw_{\rm WZW}$ can be formally performed in $(0, 1)$ superspace as Eq.\,(\ref{SS1}):
\beq
\begin{split}
\cw_{\rm sWZW}&=\frac{c}{16\pi}{\int_B\dx {\rm d}t\int {{\rm d}\theta_R}}\ \Tr_F\left(\Psi^{-1}\partial_t\Psi[\Psi^{-1}D_L\Psi, \Psi^{-1}\partial_R\Psi]\right)\nonumber\\[1mm]
&=\frac{c}{24\pi}{\int_{\varphi(B)}\!\!\!\!\!\Tr_F(\varphi^{-1}d\varphi)^3}-\frac{ic}{16\pi}\int_{\Sigma}\dx\ \Tr_F\left(\zeta_L\varphi^{-1}\partial_{R}\varphi\zeta_L\right),
\end{split}
\label{sWZW}
\eeq
where we define superfield $\Psi$, c.f. Eq.\,(\ref{fermion def}):
$$\Psi\vert_{\theta_R=0}\equiv\varphi\ \ {\rm and}\ \ \Psi^{-1}D_L\Psi\vert_{\theta_R=0}\equiv\zeta_L\,.$$
As what we mentioned, for now all fields are gauge invariant, one should not worry about anomalies for the fermionic part $\cw_{\rm sWZW}$. Overall we have a supersymmetric effective action:
\beq
S^{(0, 1)}=S_b+S^\prime_f+\cw_{\rm sWZW}
\label{eff}
\eeq\\

\noindent {\bf iii. Renormalization flow and superconformal fixed point in IR region}\\

Now we want to investigate some non-perturbative behaviors of the modified theories in deep infrared region. It is interesting to realize that the modified theory contains supersymmetric ``WZW" term with gauge invariant variable $\varphi=g\tilde h^{-1}$. We are trying to argue that, in an \emph{ad-hoc} gauge:\\
(a) for $M$ is a symmetric space, the ``WZW" action vanishes and the theory is equivalent to a bosonic sigma model with left fermions decoupled. Therefore supersymmetry should be broken in IR region;\\
(b) for $M$ is a non-symmetric homogeneous spaces with non-trivial third cohomology $H^3(M)\neq 0$, the ``WZW" term corresponds to an element in $H^3(M)$. The theory would flow to a (super)conformal fixed point in IR region.\\

To illustrate part (a), we fix the gauge on variable $\varphi$, so that
\beq
\varphi^{-1}d\varphi\in\Omega^1(M)\otimes\mf m,\ \ {\rm or\  say}\ \ \varphi^{-1}d\varphi=e^aX_a\,,
\label{Mgauge}
\eeq
where $e^a$ will be shown as vielbein $1$-forms on $M$ soon. This gauge is alway possible to choose, although $\varphi$ cannot be expressed in terms of exponential map. It is because that, if we notice $g=\varphi\tilde h$, $\varphi$ is exactly a coset representative for $M=G/H$, and thus $\varphi^{-1}d\varphi$ is a 1-form on $\T^\ast M$.

Now under this gauge, by the property of symmetric space
$$[\mf m, \mf m]\subset \mf h,$$
and the orthogonality Eq.\,(\ref{perp decomp}), one verifies that:
$$\Tr_F(\varphi^{-1}d\varphi)^3=0,\,\ \ {\rm and}\ \ \Tr_F\left(\zeta_L\varphi^{-1}\partial_{R}\varphi\zeta_L\right)=0\,,$$
for $\zeta_L=\zeta^a_LX_a$ as well. Therefore the fermion $\zeta_L$ is totally decoupled and free. Now let us turn to bosonic part, see Eq.\,(\ref{L_G/H}). We rewrite the action in the light-cone coordinate as
$$S_{M}=-\frac{1}{2\lambda^2}\int_\Sigma\dx\,\Tr_F\Big[(g^{-1}\partial_R g-A_R)(g^{-1}\partial_L g-A_L)\Big].$$
By using Eqs.\,(\ref{varphi}) and (\ref{PW}) we further express the action in terms of $\varphi$, $\tilde h$ and $A_L$:
$$S_{M}\!=\!-\frac{1}{2\lambda^2}\!\int_\Sigma\!\dx\,\Tr_F\Big[(\varphi^{-1}\partial_{R}\varphi)(\varphi^{-1}\partial_{L}\varphi)+(\varphi^{-1}\partial_{R}\varphi)(\partial_L\tilde h\tilde h^{-1})-(\varphi^{-1}\partial_{R}\varphi)(\tilde hA_L\tilde h^{-1})\Big].$$
The last two terms vanish because of orthogonality again. Therefore we finally have
\beq
S^{(0, 1)}_M=-\frac{1}{2\lambda^2}\!\int_\Sigma\!\dx\, \Tr_F\Big[(\varphi^{-1}\partial_{R}\varphi)(\varphi^{-1}\partial_{L}\varphi)+i(\zeta_L\partial_R\zeta_L)\Big].
\label{L_S}
\eeq
It is well-known that the bosonic theory is asymptotic free. In the deep infrared region, there is a mass gap generated, while the free fermions $\zeta_L$ is chiral, and thus no way to pair mass term. Thereby the supersymmetry will be broken. 

For sure when we use Eq.(\ref{PW}) to parametrize gauge fields, there are functional determinants raised,
$$\int\mathcal DA_R\mathcal DA_L=\int(\mathcal Det\nabla_R)(\mathcal Det\nabla_L)\mathcal D\tilde h\mathcal D\tilde h^\prime\,,$$
where $\nabla_{R, L}\equiv\partial_{R, L}+[A_{R, L}, \ \ ]$. The two determinants combining together is gauge invariant, and gives an additional Polyakov-Wiegmann functional \cite{Polyakov:1984,Schnitzer:1989},
$$(\mathcal Det\nabla_R)(\mathcal Det\nabla_L)=\exp(-	ic_H\cw_{\rm PW}[\tilde h\tilde h^{\prime -1}])\partial_R\partial_L\,,$$
where $c_H$ is the eigenvalue of second Casimir operator for $\mf h$ in its adjoint representation. Nevertheless, this additional term will not affect our argument above.\\

Now we are aiming to argue part (b) under the same gauge Eq.\,(\ref{Mgauge}). Since for non-symmetric homogeneous spaces, the Lie algebra structure constant $C_{abc}$ is non-zero, we will have non-vanishing WZW term and fermionic interaction, see Eq.\,(\ref{sWZW}). The WZW term
$$\frac{c}{24\pi}\Tr_F(\varphi^{-1}d\varphi)^3\sim C_{abc}e^a\wedge e^b\wedge e^c$$
is a closed and horizontal basic $3$-form, which vanished under action of $\mf h$-Lie derivative $\mathcal L_{\mf h}$. Therefore it is an element in $H^3(M)$, when $H^3(M)\neq 0, $ c.f. \cite{D'Hoker:1994} and \cite{de_Azcarraga:1998}. Combining this term with original bosonic action, see Eq.\,(\ref{L_S}), we have
$$S_{M, b}=-\frac{1}{2\lambda^2}\int_\Sigma\dx\ \Tr_F(\varphi^{-1}\partial_{R}\varphi)(\varphi^{-1}\partial_{L}\varphi)+\frac{c}{24\pi}{\int_{\varphi(B)}\!\!\!\!\!\Tr_F(\varphi^{-1}d\varphi)^3}\,.$$
By standard argument, we know, that for\,\footnote{\,For simplicity, we only assume one coupling constant, say $\lambda^2$, even though $\varrho$ may be reducible.}
\beq
\frac{\lambda^2c}{8\pi}=1
\label{scp}
\eeq
the bosonic theory will be conformal invariant. Now let us temporarily reside at this critical point, and check the fermionic action. Combining Eq.\,(\ref{Sf'}) and the fermionic part of Eq.\,(\ref{sWZW}), we get
\beq
\begin{split}
S_{M, f}&=S^\prime_f-\frac{ic}{16\pi}\int_{\Sigma}\dx\ \Tr_F\left(\zeta_L\varphi^{-1}\partial_{R}\varphi\zeta_L\right)\\[1mm]
&=-\frac{ic}{16\pi}\int_\Sigma\dx\ \Tr_F\left(\zeta_L\partial_R\zeta_L+2\zeta_L\varphi^{-1}\partial_{R}\varphi\zeta_L\right).
\end{split}
\eeq
Similar to Eq.\,(\ref{psi-zeta}), we further rotate $\zeta_L$ to define a new fermionic variable $\xi_L$ satisfying
$$\zeta_L\equiv\varphi^{-1}\xi_L\varphi\,.$$
We obtain a free fermionic action on $\xi_L$ as:
$$S_{M, f}=-\frac{ic}{16\pi}\int_\Sigma\dx\ \Tr_F\xi_L\partial_R\xi_L\,.$$
Certainly such a redefinition on chiral fermions will lead us to the Polyakov-Wiegmann functional as before, although our theory has been gauge invariant as it was modified. Such an additional functional seems only to contribute a shift to the level $c$ of the conformal theory. In sum, because of the existence of WZW term, the theory will flow to a non-trivial infrared conformal fixed point, where fermionic fields are free, while due to conformal symmetries, there is no mass gap for bosonic sector, and thus the $(0, 1)$ supersymmetry seems to hold.
\newpage

\subsection{Examples}
\label{3.3}

In this subsection, we turn to use anomaly matching condition Eq.\,(\ref{amc}) to analyze some examples.\\ 

\noindent {\bf i. Simple Lie group G}\\

Our first example is sigma models defined on simple Lie groups $G$. Although we construct sigma models on $M=G/H$ by gauging a subgroup $H$ of $G$, see Eq.\,(\ref{SS1}), Lie group $G$ itself is a symmetric space as well, i.e.
$$G\simeq G_L\times G_R/G_V\,.$$
The Lie algebra of $G_L\times G_R$ is
$$\mf g_L\oplus\mf g_R,\ \ {\rm with\ that}\ \ \mf g_L=\mf g_R=\mf g\ .$$
We label the generators $L_A\in\mf g_L$ and $R_A\in\mf g_R$, their commutators are 
$$[L_A,\ L_B]=C^C_{AB}L_C\,, \  \  \ [R_A,\ R_B]=C^C_{AB}R_C\,,\  \  \  [L_A,\ R_B]=0\,.$$
The diagonal group $G_V$ acting on $G_L\times G_R$ gives its Lie subalgebra $H_A\in \mf g_V$\,,
$$H_A=L_A+R_A\,.$$
By using Killing form with normalization
$$\Tr(L_AL_B)=-\delta_{AB},\ \ \Tr(R_AR_B)=-\delta_{AB},\ \ {\rm and}\ \ \Tr(L_AR_B)=0\,,$$
we find other generators belonging to $\mf m$\,, complimentary to $\mf h=\mf g_V$\,,
$$X_A=L_A-R_A\,,$$
and their commutator relationship given by
$$[H_A,\ H_B]=C^C_{AB}H_C\,,\ \ [H_A,\ X_B]=C^C_{AB}X_C\,,\ \ {\rm and}\ \ [X_A,\ X_B]=C^C_{AB}H_C\,.$$
Therefore $G\simeq G_L\times G_R/G_V$ is a symmetric space with isotropy representation $\varrho$\,,
$$({\rm ad}\ \mf g_L\oplus\mf g_R)\vert_{\mf g_V}={\rm ad}\ \mf g_V\oplus\varrho={\rm ad}\ \mf g_V\oplus{\rm ad}\ \mf g_V\,.$$
And we see that
$$\Tr_\varrho(H_AH_B)=-T_G\delta_{AB}=\frac{T_G}{2}\,\Tr_F(H_AH_B)\,,$$
where $T_G$ is the dual Coxeter number of Lie algebra $\mf g$. By anomaly matching condition, we know that sigma model is well-defined on simple group manifold $G$.

Another motivation for us to consider sigma model on $G$ is that: we want to argue that ease of holonomy anomalies, independence of the theory on choices of section $s$ is \emph{prior} to that of isometry anomalies. To illustrate this point, we first look at the action Eq.\,(\ref{superL_G/H}), without gauge and gaugino fields $A_\mu$ and $\chi_R$\,,
\beq
S^{(0,1)}_{G}=-\frac{1}{2\lambda^2}\int_\Sigma\dx\ \Tr(g^{-1}\partial_{L}g)(g^{-1}\partial_{R}g)+i\Tr\left(\psi_L(\partial_{R}+g^{-1}\partial_{R}g)\psi_L\right)\ .
\label{L'_G}
\eeq
For this action, in fact, we already fix a gauge, or say a section $s:U_s\subset G\rightarrow G_L\times G_R$. Near the identity of $G_L\times G_R$, one can assign coordinates $\{\phi\}\in U_s$ and use exponential map to write $s$ explicitly,
\beq
g=s(\phi)={\rm e}^{2\phi^AL_A}\ .
\label{sG}
\eeq
From the above equation, we also know that the gauge fixing is to remove degrees of freedom on $G_R$. Let us keep it in mind. In the following we will show, under this gauge fixing, there is \emph{no} isometry anomaly.

We consider isometries of the action Eq.\,(\ref{L'_G}). One can either interpret these isometries as left isometries of $G_L$ and right ones of $G_R$, or as \emph{all left} isometries acting on Eq.\,(\ref{sG}). Isometries of $G_R$, parameterized by ${\rm e}^{\epsilon^AR_A}$, acting on $s(\phi)$ from \emph{left}, break the fixed gauge,
$${\rm e}^{\epsilon^AR_A}s(\phi)={\rm e}^{\epsilon^AR_A+2\phi^AL_A}\,.$$
Therefore we need compensate it by a $G_V$-gauge transformation $h={\rm e}^{-\epsilon^AH_A}$,
$${\rm e}^{\epsilon^AR_A}s(\phi)h={\rm e}^{2\phi^AL_A}{\rm e}^{-\epsilon^AL_A}\ ,$$
where in the two equations above we used the fact that $L_A$ and $R_B$ commutes, and thus it is equivalent to a \emph{right} $G_R$ group action. Since isometries from $G_L$ need no gauge compensation, whereas isometries from $G_R$ need a \emph{constant} gauge compensation, say $h={\rm e}^{-\epsilon^AH_A}$, both $G_L$ and $G_R$ isometries do not produce isometry anomalies in the choice of section $s$. One can also confirm this statement directly from the fermionic part of action Eq.\,(\ref{L'_G}),
\beqn
&&{\rm for}\ g\rightarrow kg,\ \  \ g^{-1}\partial_{R}g\rightarrow g^{-1}\partial_{R}g\,;\nonumber\\
&&{\rm for}\ g\rightarrow g\tilde k,\ \  \ g^{-1}\partial_{R}g\rightarrow \tilde k^{-1}(g^{-1}\partial_{R}g)\tilde k\,,\nonumber
\eeqn
where $k, \tilde k\in G_L, G_R$ are constant group elements. We get the same result that $g^{-1}\partial_{R}g$ is invariant under left isometries, and tensorially transformed under right ones. Hence, after integrating out fermions from action Eq.\,(\ref{L'_G}), the fermionic effective action will not produce isometry anomalies.

From the analysis above, it seems that the theory is well-defined even with no need to add counterterms. However in what follows, we will argue that introducing counterterms as Eq.(\ref{ct}) is a must. First we notice that there is a discrete symmetry \emph{classically} held. On bosonic part of the action Eq.\,(\ref{L'_G}), we realize that
$$g\rightarrow g^{-1},\ \ S_{G, b}\rightarrow S_{G, b}\,.$$
On the other hand, the fermionic part is changed to 
$$S_{G, f}\rightarrow -\frac{i}{2\lambda^2}\int_{\Sigma}\dx\ \Tr\big[\psi_L(\partial_{R}+g\partial_{R}g^{-1})\psi_L\big]\,.$$
To get it back to $S_{G, f}$, one need rotate chiral fermions $\psi_L$ simultaneously with $g$, 
$$g\rightarrow g^{-1},\ \ {\rm and}\ \ \psi_L\rightarrow g\psi_Lg^{-1}\,.$$
Now since the transformation of chiral fermion is $x$-dependent, such a rotation will produce an integrated anomaly at quantum level, which is a WZW-like term that breaks this symmetry explicitly. Adding a WZW-like counterterm as 
Eq.\,(\ref{ct}) is exactly to offset this anomaly and keep the discrete symmetry above. So far it is still not adequate to require a counterterm, for there is no priori to admit this discrete symmetry in our theory. In fact, on the contrary, a four-dimensional sigma model describing goldstone bosons denies the symmetry $g\rightarrow g^{-1}$, but require it accompanied by parity inversion on spacetime, c.f. \cite{Witten:1983}.  

Nevertheless, in our case, the anomaly of this discrete symmetry is a signal of non-equivalence of different choices of sections, or gauge fixings. To see this, let us recall CCWZ coset construction on group $G$ manifold, i.e.\ the unitary gauge Eq.\,(\ref{unitary gauge}),
\beq
s^\prime(\phi)={\rm e}^{\phi^AX_A}={\rm e}^{\phi^AL_A-\phi^AR_A}\ .
\label{CCWZ}
\eeq
Under this gauge, we describe our theory by writing its vielbeins and connection. From Eq.\,(\ref{decomp mc1form}) and Eq.\,(\ref{pullback}), we have the pullback Maurer-Cartan 1-form
$$s^{\prime -1}ds^{\prime}={\rm e}^{-\phi^AL_A}d{\rm e}^{\phi^AL_A}+{\rm e}^{\phi^AR_A}d{\rm e}^{-\phi^AR_A}$$
for $L_A$ and $R_B$ commute. Further, because $L_A$ and $R_B$ satisfy the same commutation rules, we will have same functional form, $\theta(\phi)$ for example, for the two terms with arguments up to a minus sign, i.e.,
$$s^{\prime -1}ds^{\prime}=\theta^A(\phi)L_A+\theta^A(-\phi)R_A=\frac{1}{2}\big[\theta^A(\phi)+\theta^A(-\phi)\big]H_A+\frac{1}{2}\big[\theta^A(\phi)-\theta^A(-\phi)\big]X_A\,.$$
From it, we can read off the vielbein and connection 1-form under unitary gauge,
$$
e^{\prime A}(\phi)=\frac{1}{2}\big[\theta^A(\phi)-\theta^A(-\phi)\big],\ \ \omega^{\prime A}(\phi)=\frac{1}{2}\big[\theta^A(\phi)+\theta^A(-\phi)\big] .$$
Apparently, $e^{\prime A}(\phi)$ and $\omega^{\prime A}(\phi)$ are odd and even 1-forms separately. One can check that, with the help of the parities of $e^\prime$ and $\omega^\prime$, the theory indeed has the discrete symmetry mentioned above, which in coordinates $\phi$ and fermions $\psi_L$ is given as:
\beq
\phi\rightarrow -\phi,\ \ \psi_L\rightarrow -\psi_L\,.
\label{parity}
\eeq
Since the fermions is intact, at quantum level, this discrete symmetry is still hold. On the other hand, if we choose the section as Eq.\,(\ref{sG}), the Maurer-Cartan 1-form is
$$s^{-1}ds=\theta^A(2\phi)L_A=\frac{1}{2}\,\theta^A(2\phi)H_A+\frac{1}{2}\,\theta^A(2\phi)X_A\equiv\omega^A(\phi)H_A+e^A(\phi)X_A\,.$$
In this gauge, vielbeins and connection 1-form coincide\,\footnote{\,It should be noticed that, although they have the same form, but they follow different transformation rules, see Eq.\,(\ref{H-gauge trans rho}). This difference will not be detected by isometric transformation, for they only induce \emph{constant} gauge transformation as what we showed.} with each other, but their parities are sacrificed. 

Now we are in the situation that we do not ask for the theory to have or deny the symmetry (\ref{parity}), but rather require it to be equivalently described in different choices of sections, e.g. $s$ or $s^\prime$. We know that sections 
(\ref{sG}) and (\ref{CCWZ}) are connected by a $H$-gauge transformation,
$$s^\prime(\phi)=s(\phi){\rm e}^{-\phi^AH_A}\,.$$
Therefore the theory Eq.\,({\ref{L'_G}}) is required to be $H$-gauge invariant even it has been shown to have vanishing isometry anomalies.

Furthermore, with the counterterm (\ref{ct}) added, applying the result of Sec.\ref{3.2}, we know that the $\mathcal N=(0, 1)$ supersymmetric sigma model defined on simple Lie group $G$ is equivalent to its bosonic principal sigma model plus a free chiral fermions, which is also different from the one predicted by action Eq.\,(\ref{L_G}).\\\\

\noindent {\bf ii. Oriented real Grassmannian manifolds}\\

Our second example is oriented real Grassmannian manifolds:
$$M=\frac{{\rm SO}(p+q)}{{\rm SO}(p)\times{\rm SO}(q)}\,.$$
We have known that, for $p=1$ (or $q=1$), the manifolds is just sphere $S^{\,q}$ with vanishing isometry anomalies\cite{CCSV2}. Now we will consider the more general case by anomaly matching condition Eq.\,(\ref{amc}).

In the Grassmannian case, $G={\rm SO}(p+q)$ and $H={\rm SO}(p)\times{\rm SO}(q)$ with standard embedding. We choose generators $T_{AB}$ in the fundamental representation of Lie algebra $\mf g=\mf{so}(p+q)$ as:
$$(T_{AB})_{CD}=-\delta_{AC}\delta_{BD}+\delta_{AD}\delta_{BC}\,,\ \ {\rm with}\ A, B, C, D=1, 2, ..., p, p+1, ..., p+q\,.$$
Their  commutators are
$$[T_{AB},\ T_{CD}]=\delta_{AC}T_{BD}+\delta_{BD}T_{AC}-\delta_{AD}T_{BC}-\delta_{BC}T_{AD}\,,$$
and the normalized by Killing form is
$$\Tr(T_{AB}T_{CD})=-2(\delta_{AC}\delta_{BD}-\delta_{AD}\delta_{BC})\,.$$
For Lie subalgebra $\mf h=\mf{so}(p)\oplus\mf{so}(q)\equiv\mf h_p\oplus\mf h_q$ we label generators as
\beqn
&&H_{\mathbf i}\equiv T_{ij}\in\mf {so}(p)\,,\ \ \ {\rm for}\ i, j=1, 2,..., p\,,\nonumber\\[1mm]
&&H_{\mathbf a}\equiv T_{ab}\in\mf {so}(p)\,,\ \ \ {\rm for}\ a, b=p+1, p+2,..., p+q\,,\nonumber
\eeqn
where we use subscripts ``$\mathbf i$" and ``$\mathbf a$" to label two indices for brevity. The rest of generators forms subspace $\mf m$ complimentary to $\mf h$, where we label them as
$$X_{ia}\equiv T_{ia}\,,\ {\rm for}\ i=1,2,...,p\,;\ a=p+1, p+2,..., p+q\,.$$
Now we will investigate the isotropy representation of $\mf h$ on $\mf m$. For
$$[\mf h_p,\ \mf h_q]=0\,,$$
we have the decomposition by Eq.\,(\ref{restrict}),
\beq
({\rm ad}\ \mf g)\vert_{\mf h}=({\rm ad}\ \mf h)\oplus\varrho=({\rm ad}\ \mf h_p\oplus{\rm ad}\ \mf h_q)\oplus\mf \varrho\,.
\label{Grassman}
\eeq
Actually we only need to care about $\Tr_\varrho (H_{\mathbf i}H_{\mathbf j})$, $\Tr_\varrho (H_{\mathbf a}H_{\mathbf b})$ and $\Tr_\varrho (H_{\mathbf i}H_{\mathbf b})$. From Eq.\,(\ref{Grassman}), we have the equality
\beq
\Tr_\varrho=\Tr_{{\rm ad}\,\mf g}-\Tr_{{\rm ad}\,\mf h}\,,
\label{trace}
\eeq
while the latter two traces are easy to calculate by the commutation relationship and normalization above. After a short calculation,
\beqn
&\Tr_{{\rm ad}\,\mf g}(H_{\mathbf i}H_{\mathbf j})\!=\!-2(p\!+\!q\!-\!2)\delta_{\mathbf{ij}}\,,\ \Tr_{{\rm ad}\,\mf g}(H_{\mathbf a}H_{\mathbf b})\!=\!-2(p\!+\!q\!-\!2)\delta_{\mathbf{ab}}\,,\ \Tr_{{\rm ad}\,\mf g}(H_{\mathbf i}H_{\mathbf b})=0\,;\nonumber\\[1mm]
&\Tr_{{\rm ad}\,\mf h}(H_{\mathbf i}H_{\mathbf j})=-2(p-2)\delta_{\mathbf{ij}}\,,\ \Tr_{{\rm ad}\,\mf h}(H_{\mathbf a}H_{\mathbf b})=-2(q-2)\delta_{\mathbf{ab}}\,,\ \Tr_{{\rm ad}\,\mf h}(H_{\mathbf i}H_{\mathbf b})=0\,.\nonumber
\eeqn
Thus, we have
$$\Tr_{\varrho}(H_{\mathbf i}H_{\mathbf j})=-2q\,\delta_{\mathbf{ij}}\,,\ \Tr_{\varrho}(H_{\mathbf a}H_{\mathbf b})=-2p\,\delta_{\mathbf{ab}}\,,\ \Tr_{\varrho}(H_{\mathbf i}H_{\mathbf b})=0\,.$$
Therefore, to meet anomaly matching condition, we  have only two cases when minimal $\mathcal N=(0, 1)$ supersymmetric sigma models exist.\\[2mm]
Case 1: $p=1$, $M=S^{\,q}$\,: ~$\Tr_\varrho(H_{\mathbf a}H_{\mathbf b})=\Tr_F(H_{\mathbf a}H_{\mathbf b})$.\\[2mm]
Case 2: $p=q$, $M={\rm SO}(2p)/({\rm SO}(p)\times {\rm SO}(p))$\,: ~$\Tr_\varrho(H_{\mathbf i}H_{\mathbf j})=p\,\Tr_F(H_{\mathbf i}H_{\mathbf j})\,.$\\[2mm]
The result on case 2 should have no further difficulty to be generalized to the case that $H$ contains more than two identical factors, $H\simeq H_1\times H_2\times\cdots\times H_n$.\\ 

For the anomaly on oriented real Grassmannian manifolds $M$, there is also another interesting observation that helps verify our anomaly matching condition. Instead of constructing sigma models on ${\rm SO}(p+q)$ followed by gauging its subgroup ${\rm SO}(p)\times {\rm SO}(q)$, one can consider another fiberation,
$$\xymatrix{{\rm SO}(p) \ar[r]^{\!\!\!\!\!\! i} & V_q(\mathbb R^{p+q})\ar[r]^{\ \ \ \ \pi} & M}\,,$$
where  $V_q(\mathbb R^{p+q})\simeq{\rm SO}(p+q)/{\rm SO}(q)$ is the Stiefel manifold which is the set of all orthonormal $q$-frames in $\mathbb R^{p+q}$. Sigma models built on Stiefel manifold is always well-defined which we will show in our next example. Here let us just assume it and consider how a real Grassmannian sigma model can be constructed in the fiberation above.

In \cite{CCSV2} we introduced a dual formalism for ${\rm O}(N)$ model. With a little modification, we can work out the $\mathcal N=(0, 1)$ supersymmetric action on Grassmannian manifold,
\beq
\begin{split}
&S_M=\frac{1}{2g_0^2}\int{\dx \,\rm Tr}\left((\nabla_R n)^T\nabla_L n+i\psi_L^T\nabla_R\psi_L\right),\\[1mm]
&(n^{T})^a_{\ \alpha} n^{\alpha}_{\ b}=\delta^{a}_{\ b}\,,\qquad (n^{T})^{a}_{\ \alpha}\psi^\alpha_{Lb}=0\,,
\end{split}
\eeq
where $n^\alpha_{\ a}$  and $\psi_{La}^\alpha$ ($ \alpha=1,2,...,p+q$), ($a=1,2,...,p$),
are real bosonic fields and their chiral fermions partners, and the covariant derivative $\nabla$ is defined as
$$(\nabla_{R, L} n)^{\alpha}_{\ a}=\partial_{R, L} n^{\alpha}_{\ a}-n^{\alpha}_{\ b}A^b_{R, L a}\,.$$ 
The model is obtained by gauging the color symmetries, i.e. those on indexes ``$a, b...$", of the action on Stiefel manifolds. Thus action on Stiefel manifolds is obtained by removing gauge fields, and loosing the constraints on $n$ and $\psi_L$ as
$$n^{T}\psi_{L}+\psi^T_Ln=0\,.$$
In standard $(0, 1)$ superspace construction, one can introduce a super Lagrange multiplier $\Lambda^a_{R\,b}$,
$$\Lambda^a_{R\,b}=\lambda^a_{R\,b}+\theta_R\sigma^a_{\ b}\ ,$$
with indexes $a, b$ symmetrized. Thus, the super-constraint term is
$$S_c=\int\dx\int{\rm d}\theta_R\Tr (\Lambda_R N^TN)\,,$$
where
$$N=n+i\theta_R\psi_L$$
is the superfield version of fields $n$ and $\psi_L$. The sigma model on Stiefel manifold is given by
$$S_V=\!\int\!\dx\!\int\!{\rm d}\theta_R\Big[-\frac{1}{2g_0^2}\,\Tr\left((D_LN)^T\partial_{RR}N\right)+\Tr (\Lambda_R N^TN)\Big].$$
Correspondingly, after gauging its ``color" symmetry ${\rm SO}(p)$, we obtain sigma model on Grassmannian manifold $M$,
$$S_M=\!\int\!\dx\!\int\!{\rm d}\theta_R\Big[-\frac{1}{2g_0^2}\Tr\left((D_LN-N\mathcal V_L)^T(\partial_{RR}N-N\mathcal V_{RR})\right)+\Tr (\Lambda_R N^TN)\Big],$$
where supergauge multiplets $\mathcal V_{L, RR}$ were introduced in Eq.\,(\ref{supergauge}).

In fact the above two sigma models can be obtained by considering the (classically) low energy limit of $\mathcal N=(0, 1)$ two-dimensional gauge theories and Yukawa theories respectively. We build the Yukawa theories as
\beq
\begin{split}
S_Y&=\!\int\!\dx \!\int\!{\rm d}\theta_R\Big[-\frac{1}{2g_0^2}\Tr\left((D_LN)^T\partial_{RR}N\right)+\Tr (\Lambda_R N^TN)\Big]\\[1mm]
&+\!\int\!\dx\!\int\!{\rm d}\theta_R\Big[-\frac{1}{2\lambda_0^2}\Tr(\Lambda_R^TD_L\Lambda_R)\Big]. 
\end{split} 
\eeq
It is noticed that the coupling constant $\lambda_0$ has mass dimension for $\Lambda_R$ has mass dimension $3/2$. In the low energy limit, we put $\lambda_0\rightarrow\infty$, and obtain the action $S_V$. In the sense we can interpret the UV completion of the sigma model on Stiefel manifold is a Yukawa theory, although the sigma model itself can be considered as a renormalizable theory in two-dimensions.

Similarly, let us find the UV completion of $S_M$ by gauging the Yukawa theory $S_Y$ and adding gauge sectors. Noticing that the Yukawa interaction $S_c$ is gauge invariant, we have
\beq
\begin{split}
S_{Y+\mathcal V}&=\!\int\!\dx\!\int\!{\rm d}\theta_R\Big[-\frac{1}{2g_0^2}\Tr\left((D_LN-N\mathcal V_L)^T(\partial_{RR}N-N\mathcal V_{RR})\right)+\Tr (\Lambda_R N^TN)\Big]\\[1mm]
&+\!\int\!\dx\!\int\!{\rm d}\theta_R\Big[-\frac{1}{2\lambda_0^2}\Tr\left(\Lambda_R^T(D_L\Lambda_R+[\mathcal V_L,\ \Lambda_R])\right)\Big]\\[1mm]
&+ \!\int\!\dx\!\int\!{\rm d}\theta_R\Big[-\frac{1}{4e_0^2}\Tr\left(W_R(D_LW_R+[\mathcal V_L,\ W_R])\right)\Big],
\end{split}
\eeq
where $W_R$ is field strength of gauge potential $\mathcal V_{L, RR}$,
$$W_R\equiv[D_L+\mathcal V_L\,,\ \partial_{RR}+\mathcal V_{RR}\,]\,.$$
Couplings $e_0$ and $\lambda_0$ are of the same nonvanishing dimensionality, so in a low energy limit the last two terms fade away, and we obtain the sigma model $S_M$ on Grassmannian $M$.

Now due to the observation of t'Hooft's consistency condition, we should expect that $S_{Y+\mathcal V}$ and $S_{M}$ produce same anomalies or be anomaly-free. Therefore we focus ourselves on the gauge fields and bi-fermions interactions of action $S_{Y+\mathcal V}$ and calculate their anomalies. The relevant part of the Lagrangian is
\beqn
{\mathcal L}_{\rm f.A.f}&=&\frac{i}{2g_0^2}(\psi_L^T)^{a}_{\ \alpha}(\partial_R\psi_L-\psi_LA_R)^\alpha_{\ a}+\frac{i}{2\lambda^2_0}\lambda^a_{R\,b}(\partial_L\lambda^b_{R\,a}+[A_L, \lambda_R]^b_{\ a})\nonumber\\
&+&\frac{i}{2e^2_0}\chi_{R\,{\mathbf i}}(\partial_L\chi^{\mathbf i}_{R}+A^{\mathbf j}_LC^{\mathbf i}_{\mathbf {jk}}\chi^{\mathbf k}_{R})\ .
\eeqn
To see if there are gauge-anomalies produced, we need to consider a vector rotation and compare the gauge anomalies from left and right fermions. For the right, since gauge fields are in the fundamental representation and we also need sum up flavors, say $\alpha$ indexes, we finally have:
$$\mathcal A_R\sim (p+q)\Tr_F(H_{\mathbf i}H_{\mathbf j})=-2(p+q)\delta_{\mathbf {ij}}\ .$$
On the other hand, gauge fields interacting with gauginos are in adjoint representation of ${\rm SO}(p)$, and with $\lambda_{R}$ are in the fundamental representation. We have:
$$\mathcal A_L\sim\Tr_{\rm ad}(H_{\mathbf i} H_{\mathbf j})+(p+2)\Tr_F(H_{\mathbf i} H_{\mathbf j})=-4p\,\delta_{\mathbf{ij}}$$ 
Therefore gauge anomaly vanishes only when $p=q$ consistent with the result we obtained for the sigma model. 

This observation on the correspondence of anomalies in two-dimensional gauge theories and sigma models could be useful for
 considerations theories in deep infrared region. We made some predictions in Sec.\,\ref{3.2}. We expect to verify them in anomaly-free gauge theories, by considering Large $N$-expansion as well. We will present this work somewhere else in the future.\\\\

\noindent {\bf iii. \boldmath{$\left(G\times{\rm U}^r(1)\right)/H$} for group \boldmath{$G$} semi-simple and \boldmath{$H$} simple}\\ 

Our third example is inspired by Eq.\,(\ref{trace}), by which we will show that for homogeneous space $M\!=\!\left(G\times{\rm U}^r(1)\right)\!/H$ with $G$ semi-simple and $H$ simple, the anomaly matching condition Eq.\,(\ref{amc}) will be satisfied. Therefore, there always exists minimal $N=(0, 1)$ supersymmetric sigma model on them. 

The proof of the above statement is quite transparent when both $G$ and $H$ are simple groups. Since $G$ and $H$ are simple, their Lie algebras $\mf g$ and $\mf h$ will be simple, and, thus, contain no non-trivial ideals. Therefore, their adjoint representation ${\rm ad}\,\mf g$ and ${\rm ad}\,\mf h$ are irreducible respectively. By choosing an appropriate basis, generators $H_i\in \mf h$ will satisfy to the following relations,
\beq
\Tr_{{\rm ad}\,\mf g}(H_iH_j)\!=\!-T_G\delta_{ij}\,,\ \ \Tr_{{\rm ad}\,\mf h}(H_iH_j)\!=\!-T_H\delta_{ij}\,; \ \ \Tr_{F(\mf g\vert_{\mf h})}(H_iH_j)\!=\!-\delta_{ij}\,,
\label{g/h}
\eeq
where $T_G$ and $T_H$ are the dual Coxeter numbers of $\mf g$ and $\mf h$ respectively. Combining 
Eq.\,(\ref{restrict}) and Eq.\,(\ref{trace}), we have
\beq
\Tr_\varrho(H_iH_j)=-(T_G-T_H)\delta_{ij}=(T_G-T_H)\Tr_F(H_iH_j)\ .
\label{CG-CH}
\eeq

Now we can easily improve the above result. For $G$ is semi-simple, the only difference is that we have distinguished normalization for its each simple factor. By assigning independent coupling constants for each simple factors $G_\alpha\subset G$, we can still normalize,
$$\Tr_{F(\mf g\vert_{\mf h})}(H_iH_j)=-\delta_{ij}\ ,$$
as our convention. For adjoint representation of $\mf g$, we have
$${\rm ad}\,\mf g=\oplus_\alpha{\rm ad}\,\mf g_\alpha\,.$$
Therefore, the first relation in Eq.\,(\ref{g/h}) turns out to 
$$\Tr_{{\rm ad}\,\mf g}(H_iH_j)=-\Big(\sum_\alpha T_{G_\alpha}\Big)\delta_{ij}\,.$$
On the other hand, since $H$ is simple as before, the second relation of Eq.\,(\ref{g/h}) holds. Therefore
$$\Tr_\varrho(H_iH_j)=\bigg(\Big(\sum_\alpha T_{G_\alpha}\Big)-T_H\bigg)\Tr_F(H_iH_j)\,.$$
At last, the subgroup $H$ contains no $U(1)$ factors, so  the above result is thus unchanged.\\

Now let us apply this result to some typical examples. We simply enumerate some classical homogeneous (symmetric) space, satisfying the above condition, on which minimal $\mathcal N=(0, 1)$ supersymmetric sigma model can be constructed.\\

1. $G\times{\rm U}^r(1)$ with $G$ semi-simple: For those chiral fermions on $\rm U(1)$ are free;\\

2. Real, complex and symplectic Stiefel manifolds: 
$${\rm SO}(p+q)/{\rm SO}(p),\ \ {\rm SU}(p+q)/{\rm SU}(p),\ \ {\rm and}\ \ {\rm Sp}(p+q)/{\rm Sp}(p)\ ;$$

3. ${\rm SU}(n)/{\rm SO}(n)$, ${\rm SU}(2n)/{\rm Sp}(n)$, and ${\rm SO}(2n)/{\rm Sp}(n)$: All are symmetric spaces.\\

From the argument above, we see that the condition $H$ is simple is crucial. In fact, as we mentioned earlier, the anomaly matching condition is actually topological. In next subsection, we will show, for example when $H$ is simple, the first Pontryagin class $p_1(M)$ will always vanish.\\

\noindent {\bf iv. \boldmath{$H$} containing \boldmath{${\rm U}(1)$} factors}\\ 

Before proceeding to give a topological (characteristic class) explanation on ano\-maly matching condition Eq.\,(\ref{amc}), we want to consider another type of homogeneous space where the subgroup $H$ in turn contains $\rm U(1)$ factors. We are motivated by realizing that, when $H$ contains $\rm U(1)$ factors, the homogeneous spaces will have complex structure, and thus $\mathcal N=(0, 1)$ supersymmetry will be enhanced to $\mathcal N=(0, 2)$. Unfortunately, however, we will see soon that many of these sigma models are suffered from non-removable anomalies and thus do not exist.\\

First we want to make some clarification on the method used in Sec.\,\ref{3.1} to derive the anomaly matching condition. Although the Polyakov-Wiegmann functional, see Eq.\,(\ref{S_f2}), was given in the context of non-Abelian gauge theories, it is also true when we have some Abelian $\rm U(1)$ gauge fields. For these Abelian gauge fields, labeled by $B^i_{R, L}T_i$ for example, $T_i\subset \mf h$ commute with all other generators in $\mf h$, and forms non-trivial center of Lie algebra $\mf h$. Therefore, in the fermionic anomalous effective action Eq.\,(\ref{S_f2}), there is no 
WZW-like terms for them, but only the second one exists, i.e.,
$$\cw_{\rm anom.}=-\frac{1}{24\pi}{\int_{\tilde h(B)}\!\!\!\!\!\Tr_\varrho(\tilde h^{-1}d\tilde h)^3}+\frac{1}{16\pi}\int_\Sigma\dx\ \Tr_\varrho \omega_R(\tilde h^{-1}\partial_L\tilde h+\partial_Lu^iT_i)\,,$$
where $\tilde h$ as before parameterize those non-Abelian gauge fields $A_R$, while $u^i$ for $B^i_R$ satisfies
$$B^i_RT_i=\partial_R u^iT_i\ .$$
Meanwhile, the counterterm (\ref{ct}), which we are able to add, will be also transformed under Abelian gauge rotation. Therefore, the anomaly matching condition will be the same as before.

Nevertheless it is because of the discrepancy above, we will show in the following that the anomaly matching condition can never be fulfilled for $H=H^\prime\times{\rm U}^r(1)$ with $H^\prime$ is semi-simple. Therefore lots of minimal $N=(0, 2)$ supersymmetric sigma models, for example complex Grassmannian manifolds ${\rm U}(p+q)/({\rm U}(p)\times{\rm U}(q))$(except for ${\rm CP}^1$), are ruled out. 

The proof is quite straightforward. With no loss of generality, let us only consider $H=H^\prime\!\times\!{\rm U(1)}$ with that both $G$ and $H^\prime$ are simple Lie groups. From a previous result, Eq.\,(\ref{CG-CH}), we see that, for $H^\prime_{i, j}\in \mf h^\prime$
$$\Tr_\varrho(H^\prime_iH^\prime_j)=(T_G-T_{H^\prime})\Tr_F(H_iH_j)\,,$$
for $[H^\prime_i\,,\, T]=0$ and has no contribution to above equation, where $T$ is the generator of $\rm U(1)$. For the same reason, 
$$\Tr_\varrho T^2=\Tr_{\rm ad\,\mf g}T^2-\Tr_{\rm ad\,\mf h}T^2=\Tr_{\rm ad\,\mf g}T^2=T_G\Tr_{F}T^2\ .$$
The anomaly matching condition is, thus, not satisfied. This finishes our proof.

A corollary we can obtain is that, for homogeneous spaces $M=G/T^{r}$ with $T={\rm U(1)}$ the torus group, anomaly matching condition Eq.\,(\ref{amc}) is satisfied. Therefore minimal $\mathcal N=(0, 2)$ supersymmetric sigma models on $G/T^r$ can be well-defined.

\subsection{Topological origin of anomaly cancellation}
\label{3.4}

In this subsection, we will establish a relation between the (local) anomaly matching condition and the global topological property of homogeneous spaces $M=G/H$. More concretely, we will show that the anomaly matching condition 
Eq.\,(\ref{amc}) will be satisfied if and only if the first Pontryagin class on $M$ vanishes, i.e. $p_1(M)=0$, which thereby agrees with Moore-Nelson's constraint in case of homogeneous spaces\cite{Moore:1984ws}. 

The main argument is based on a proposition in \cite{Singhof:1982}, see Prop.\,(3.2), and a main theorem due to Borel and Hirzebruch, see Theorem 10.7 in \cite{Borel:1958}. We here only rephrase the result in terms of anomaly matching condition Eq.\,(\ref{amc}). The idea can be intuitively interpreted by Eq.\,(\ref{trace}),
$$\Tr_\varrho(H_iH_j)=\Tr_{{\rm ad}\,\mf g}(H_iH_j)-\Tr_{{\rm ad}\,\mf h}(H_iH_j)\,.$$
Since we always can, by rescaling coupling constants, require the equality
$$\Tr_{{\rm ad}\,\mf g}(H_iH_j)=c^\prime\,\Tr_F(H_iH_j)\,,$$
the anomaly matching condition is thus equivalent to
\beq
\Tr_{{\rm ad}\,\mf h}(H_iH_j)=c^{\prime\prime}\,\Tr_F(H_iH_j)\sim\Tr_{{\rm ad}\,\mf g}(H_iH_j)\,,
\label{BHBG}
\eeq
where $c^\prime$ and $c^{\prime\prime}$ are some constants. Now that we evaluate the traces in $\mf h$ and $\mf g$-adjoint representations, one can express them by means of group theoretical invariants, say symmetric functions of roots for $\mf h$ and $\mf g$ respectively. These symmetric functions are directly related to characteristic classes of $M$. We will explain that, when Eq.\,(\ref{BHBG}) is satisfied, the first Pontryagin class $p_1(M)=0$\,.

In the following, $G$ is assumed to be a compact connected Lie group, $H$ a closed subgroup of $G$, and $T\!\subset \!G$ and $S\subset H$ are maximal tori of $G$ and $H$ respectively, chosen properly so that $S\!\subset \!T$. Let $\mf s\!\subset \! \mf t$ be corresponding Lie algebras of $S$ and $T$, say the Cartan algebra of $H$ and $G$. We further set $\{\beta_1,...,\beta_s\}\subset \mf s^\ast$ and $\{\alpha_1,...,\alpha_t\}\subset \mf t^\ast$ as positive roots in respect to $H$ and $G$, and arrange them satisfying 
$$\beta_1=\alpha_1\vert_{\mf s}\,,\ \beta_2=\alpha_2\vert_{\mf s}\,,\ldots,\ \beta_s=\alpha_s\vert_{\mf s}\,,$$
and define
$$\gamma_1\equiv\alpha_{s+1}\vert_{\mf s}\,,\ \gamma_2\equiv\alpha_{s+2}\vert_{\mf s}\,,\ldots,\ \gamma_{t-s}\equiv\alpha_t\vert_{\mf s}\,,$$
which are called the roots complimentary to $H$. With the help of roots, one can rewrite the traces in Eq.\,(\ref{BHBG}) on generators $S_i\in S\subset H$ and $T_i\in T\subset G$ as
$$\Tr_{{\rm ad}\,\mf h}(S_iS_j)=\sum_{b=1}^s\beta_b(S_i)\beta_b(S_j),\ \ {\rm and}\ \ \Tr_{{\rm ad}\,\mf g}(T_iT_j)=\sum_{a=1}^t\alpha_a(T_i)\alpha_a(T_j)\,.$$
Therefore the trace operator can be expressed in terms of quadratic symmetric polynomials on $\sum\alpha^2_a$ or $\sum\beta^2_b$ on Cartan algebra $\mf s$ and $\mf t$ respectively.  Actually it is sufficient to focus ourselves only on the Cartan algebra $\mf s$ and $\mf t$. It is because that, in our case, Lie algebra $\mf h$ and $\mf g$ can be always regarded as direct sum of several simple algebras and $\mf u(1)$ factors. For each simple factor, with proper basis (Cartan-Weyl basis for example), evaluation of trace on Cartan algebra and other generators can be normalized same, but \emph{not} same among different simple factors.

On the other hand, one can identify $\{\gamma_c\}$, the set of complimentary to $H$ roots,  with $H^1(S; Z)$, the first cohomology class of tori $S$,  since they are integral functionals in ${\rm Hom}(\pi_1(S), Z)={\rm Hom}(H_1(S; Z), Z)$. The $H^1(S; Z)$ can be further identified with $H^2(BS; Z)$ via transgression, where $BS$ is the classifying space of tori $S$. Therefore, complimentary roots $\{\gamma_c\}$ will be considered as elements in $H^2(BS; Z)$. In what follows we will only work under real cohomology which will considerably simplifies our argument. First, the inclusion map
$$i: S\rightarrow H$$
induces an isomorphism $i^\ast$ on the cohomology rings of $BH$ and $BS$\,,
\beq
i^\ast: H^\ast(BH; R)\simeq H^\ast(BS; R)^{W(S, H)}\ ,
\label{i iso}
\eeq
where $H^\ast(BS; R)^{W(S, H)}$ denotes those elements invariant under Weyl group $W(S, H)$. Secondly, 
from the universal fiberation:
$$\xymatrix{G \ar[r]^i & EG \ar[r]^\pi & BG}\,,$$
we have the fiberation by module $H$,
$$\xymatrix{G/H \ar[r]^j & BH \ar[r]^q & BG}\,.$$
It induces the exact cohomology classes chain
$$\xymatrix{H^\ast(BG, R) \ar[r]^{q^\ast} & H^\ast(BH; R) \ar[r]^{j^\ast} & H^\ast(G/H; R)}\,.$$
Since $\T(G/H)$ is the vector bundle associated to $H$-principle bundle, see Eq.\,(\ref{AVH}), the total Pontryagin classes $p(G/H)$ are pullback elements from some universal elements $a\in H^*(BH; R)$\,,
$$p(G/H)=j^\ast(a)\,.$$
Now, with the identification in Eq.\,(\ref{i iso}), we can express elements $a\in H^\ast(BH; R)$ in terms of symmetric functions of complimentary roots $\gamma_c$ in $H^\ast(BS; R)^{W(S, H)}$\,,
$$a=\prod_{c=1}^{t-s}(1+\gamma_c^2)\ .$$
Specific to the first Pontryagin class $p_1(G/H)$, we have
$$p_1(G/H)=j^\ast\Big(\sum_{c=1}^{t-s}\gamma_c^2\Big)\,.$$
From the exact sequence a vanishing $p_1$ is equivalent to 
$$\sum_{c=1}^{t-s}\gamma_c^2\in {\rm Im}\,q^\ast\,.$$
At last, it is noticed that
$$\sum_{c=1}^{t-s}\gamma_c^2=\sum_{a=1}^{t}\alpha^2_a\vert_{\mf s}-\sum_{b=1}^s\beta^2_b\,,$$
while similar to Eq.\,(\ref{i iso}), we have an isomorphism on $BG$ and $BT$\,,
$$H^\ast(BG; R)\simeq H^\ast(BT; R)^{W(T, G)}\ .$$ 
Since $\sum\alpha^2_a$ is always in $H^\ast(BG; R)$, the condition $p_1(G/H)=0$, or say $\sum\gamma_c^2\in {\rm Im}\ q^\ast$, is equivalent to requirement
$\sum\beta^2_b\in {\rm Im}\,q^\ast$. It is just the anomaly matching condition ($\ref{BHBG}$).\\

\section{The determinant line bundle of homogeneous space sigma models}
\label{4}

The aim of this section is twofold. On the one hand, we would like to see in the nonlinear formulation, how much our understanding of gauge anomalies can benefit us in understanding anomalies in a pure geometric model. Isometries on Riemannian manifolds come in various cases, where some gauge formulation is far from reaching. Still one would like to understand, for example, the relation between chiral anomalies, isometry anomalies and topological anomalies. On the other hand, so far as sigma models on homogeneous spaces are concerned, we would like to see how could the gauge-like holonomy anomalies rise in a view toward determinant line bundle of certain Dirac operators parameterized by the space of bosonic field. The hope is to gain a full picture that touches each of the four corners: local vs global, gauge vs nonlinearity. A context like this can be useful in exploring interesting mathematical structures that closely tied up to each corners.

\subsection{A digression on K\"ahler sigma model anomaly\\ in Fujikawa's method}
\label{4.1}

Here we shall look at the issue of local anomaly in geometric formulation. Isometries in our system form a subset of the diffeomorphism group of the target manifold, which is accomplished via field-redefinition alone. We would like to explore, whether such symmetries remain in the quantized system, and what does the anomaly imply. Since we shall not be dealing with unphysical degree of freedom, this is similar to the case of axial anomaly and thus Fujikawa's method can be generalized to our current situation. 

We first clarify the types of manipulations we shall use in the discussion. Consider a vector field on $X$, which locally is given by $V=K^i(x)\partial_{i}$, where $x^i$s are the local coordinates on $X$. There are two possible manipulations that can be induced by $V$ --- namely the field redefinition and the infinitesimal diffeomorphism. The former is via $$\phi^i\to \phi^i+\epsilon K^i(\phi)$$ where $\phi\in C^\infty(\Sigma, X)$ is a bosonic field. Since this does not correspond to any symmetry in the action, this shall generally change the interaction. However, the field redefinition is a valid manipulation in field theories which should not cause any observational phenomena. The reason is that one can always get a contribution from the Jacobian of the path integral measure to overcome the change. The diffeomorphism transformation, on the other hand, is the aforementioned field redefinition together with the induced tensorial transformation for all geometric quantities. For example, under such transformation, the metric tensor transforms according to 
$$g_{ij}(\phi)\to g_{ij}(\phi)\,\frac{\partial \phi^i}{\partial \phi^k+\epsilon K^k(\phi)}\,\frac{\partial \phi^j}{\partial \phi^l+\epsilon K^l(\phi)}\,,$$ and $\partial \phi^i$ transforms as a tangent vector. This definitely preserves the Lagrangian at the classical level. But in field theory language, when one interprets $g_{ij}$ as a function of the field $\phi$, there will also be an accompanying transformation for the ``coupling constants" of $\phi$ in $g_{ij}$. To make sense of those, one can view those (infinite number of) constants as background fields damped at classical values. The path integral measure would need further justification. This, however, is not the case that we are interested in. 

The isometry symmetries are a subset in both classes. Defined solely by field redefinition, it satisfies the property that 
\beq
g_{ij}(\phi)\,\frac{\partial \phi^i}{\partial \phi^k+\epsilon K^k(\phi)}\,\frac{\partial \phi^j}{\partial \phi^l+\epsilon K^l(\phi)} = g_{kl}(\phi+\epsilon K(\phi))+O(\epsilon^2)
\eeq
and hence preserves the Lagrangian at the classical level. The same is true for the quantum bosonic model, and this is a pure consequence of the property of field redefinition. Indeed, we are forced to have that the Jacobian from path integral measure cancels the anomalous effective action. But a perturbative calculation shows that the effective action respects the isometry, thus forcing the path integral measure to respect the same symmetry up to an overall factor. Indeed, one can see this explicitly by writing down explicitly the measure, where we have used a standard volume form $[D\phi]$ on $X$ associated to the metric $g$,
\beq
[D\phi]= \sqrt{{\rm det}\, g_{ij}} \,d\phi^1\!\wedge\cdots d\phi^n \,.
\eeq

If our model is coupled with chiral fermions, the path integral measure might not preserve such symmetries, and if this is true, nor shall the effective action after integrating out the fermions. This is the anomaly that we are interested in. 

In supersymmetric models with target manifold $X$, fermions take value in the tangent bundle $TX$. To build the path integral measure, one has to contract the indexes on $TX$ using a standard volume form. Together with the contribution from the bosonic part, we have that
\beq
[D\psi] = \frac1{\sqrt{{\rm det}g_{ij}}}\,d\psi_L^1\cdots d\psi_L^n d\psi_R^1\cdots d\psi_R^n\,.
\eeq
Note that $\psi_R$ are decoupled from our system, and we write them down to show the comparison between our case, and the nonchiral case. Also to use Fujikawa's method, it is important to have Dirac fermions. Now we perform the isometry transformation induced by the Killing vector field $K_A$, where the index $A$ labels isometries:
\beq
\begin{split}
\phi^i (x)&\to \phi^i(y)+\epsilon^A \int \dx \,K_A^i[\phi(x)]\delta(x-y)\,,\\[1mm]
\psi_L^i (x)&\to\psi_L^i(y)+\epsilon^A\int \dx \,\partial_j K_A^i[\phi(x)]\delta(x-y)\psi^j_L(y)\,.
\end{split}
\eeq
Note that the transformation is linear with respect to the fermionic degrees of freedom. Indeed, we can learn from the case of chiral anomaly that, as far as only the local anomalies are concerned, it is really the phase factor of such transformation that matters. 

Let us suppose we have Weyl fermions. Also from here to the end of this section, we shall assume the target manifold to be K\"ahler, to get the most elegant result. Write explicitly the path integral measure as
\beq
[D\psi] = \frac1{\sqrt{{\rm det}\,G}}\, d\bar{\psi}^{\bar{1}} d\psi^{1}\cdots d\bar{\psi}^{\bar{n}} d\psi^{n}\,,
\eeq
where each $\psi$ has two components $\psi_L$ and $\psi_R$. The metric $G$ expanded in basis of $P_L\psi, P_R\psi, \bar{\psi}^{\bar{j}}P_L, \bar{\psi}^{\bar{j}}P_R$ is given by Table.\,\ref{Gexpr}. Hence \beq
{\rm det}\,G = {\rm det}\,g_{i\bar j}\,.
\eeq
\begin{table}
\begin{center}
\begin{tabular}{c|cc}
& $P_L\psi^i$ & $P_R\psi^i$\\ \hline\\[-3mm]
$\bar{\psi}^{\bar{j}}P_L$ & $0$ & $\delta_{i\bar j}$\\[1mm]
$\bar{\psi}^{\bar{j}}P_R$ & $g_{i\bar j}$ & $0$\\[-4mm]
\end{tabular}
\end{center}
\label{Gexpr}
\caption{\small The metric used in fermion path integral measure in curved indices.}
\end{table}
Now under the transformation
\beq
\begin{split}
\psi^i(x)\to &\,\psi^i(y)\!+\!\epsilon^A \mathfrak{Re}(\partial_j K_A^i[\phi(y)]) P_L\psi^j(y)\!+\!i\epsilon^A \mathfrak{Im}(\partial_j K_A^i[\phi(y)])P_L\psi^j(y)\,,\\[1mm]
\bar{\psi}^{\bar i}(x)\to &\,\bar{\psi}^{\bar i}(y) + \epsilon^A \mathfrak{Re}(\partial_{\bar j} \bar{K}_A^{\bar i}[\bar{\phi}(y)])\bar{\psi}^{\bar j}(y)P_R+ i\epsilon^A\mathfrak{Im}(\partial_{\bar j} \bar{K}_A^{\bar i}[\bar{\phi}(y)])\bar{\psi}^{\bar j}(y)P_R\\[1mm]
= &\,\bar{\psi}^{i}(y) + \epsilon^A \mathfrak{Re}(\partial_j K_A^i[\phi(y)])\bar{\psi}^{\bar j}(y)P_R- i\epsilon^A\mathfrak{Im}(\partial_j K_A^i[\phi(y)])\bar{\psi}^{\bar j}(y)P_R \,.
\end{split}
\eeq
Recall that the Jacobian, as in the pure bosonic case, has only nontrivial real part, which cancels the change of $d\psi$ and $d\bar{\psi}$ induced by $\mathfrak{Re}(\partial_j K_A^i[\phi(y)])$. But the transformation induced by $\mathfrak{Im}(\partial_j K_A^i[\phi(y)])$ is anomalous. The situation here is precisely the same as in case of chiral anomaly, and for the time being, we take the bosonic degrees of freedom to be external, or classical. 

In Fujikawa's method, infinitesimal isometry transformation gives the following extra factor for the fermion integral measure: 
\beq
\delta_{\epsilon^A}\left({\rm det}\, i\mathcal D\right)[\phi, \bar\phi] = {\rm exp}\Big(\!-i\epsilon^A\! \int\! \dx\,{\rm Tr}[\mathfrak{Im}(\partial_j K_A^i[\phi(x)]) \gamma_5] \Big)\,,
\label{measureev}
\eeq
where the trace is taken over the basis from the right eigenstates of the Dirac operator \beq i \mathcal D_{i\bar j}\equiv i\left(g_{i\bar j}\slashed\partial + g_{j\bar j}\Gamma^j_{ik}\slashed{\partial}\phi^k\right)P_L+ i\slashed{\partial}\delta_{i\bar j}P_R = i\slashed{\mathcal D}P_L +i\slashed{\partial}P_R
\eeq 
and its left eigenstates. Evaluation of Eq.\,(\ref{measureev}) is in general hard, due to the nonflatness of $g_{i\bar j}$ and the bosonic degree of freedom. But for the result in 2d, as an analog to the gauge theory case  \cite{AlvarezGaume:1983cs}, we obtain that, up to the lowest order in external fields, 
\beq
\delta_\epsilon i\Gamma_{\rm eff}[\phi] = \frac{i\epsilon_A}{4\pi}\int \mathfrak{Im}(\partial_k K_A^l)\mathcal R_{i\bar j l}^{\,\,\,\,\,\,\,\,k} d\phi^i\wedge d\bar{\phi}^{\bar j} +{\rm higher\, order\, terms\, in \,\Gamma^i_{jk}}\, .
\label{nlanoexpr}
\eeq
Indeed, we only need the leading term from $\bar\partial_{\bar j}\Gamma_{i l}^{\,\,\,\,\,k}$, which is also, up to a sign, the leading term of the curvature tensor $\mathcal R_{i\bar j k\bar l}$. Note that there is a special feature of nonabelian anomaly (or, correspondingly, the linear isometry anomaly)--- if one only cares about the lowest order in the ``gauge" field $A$ (or, correspondingly, the Christoffels $\Gamma$), then the contribution for nonabelian anomaly is, up to a constant factor, the same as the abelian anomaly \cite{Weinberg, AlvarezGaume:1983cs}. This shows why the anomaly diagram in perturbative calculation looks similar to the one involved in axial anomaly. The constant factor, in $2n$ dimensional spacetime, might depend on the kinematics of the $(n+1)$-gon Feynman diagrams. But in 2d case this is extremely simple. To determine the full structure of such anomaly, one can either do a thorough calculation of 
Eq.\,(\ref{measureev}), or use an argument like Wess-Zumino consistency condition as mentioned in \cite{AlvarezGaume:1983cs}. 

Now we calculate the explicit form of Eq.\,(\ref{nlanoexpr}) using the second method. Recall that the abelian anomaly for a nonlinear sigma model over a K\"ahler target manifold in 4d is given by \beq
{\rm Tr}\left[\mathcal R^2\right]= ( \mathcal R_{i\bar j \,\,\,\,\,\,n}^{\,\,\,\,\,\,m}\mathcal R_{k\bar l \,\,\,\,\,\, m}^{\,\,\,\,\,\, n}+
\mathcal R_{i\bar j \bar m}^{\,\,\,\,\,\,\,\,\bar n}\mathcal R_{k\bar l \bar n}^{\,\,\,\,\,\,\,\,\bar m}
)d\phi^i\wedge d\bar{\phi}^{\bar j}\wedge d\phi^k\wedge d\bar{\phi}^{\bar l}\,.
\eeq 
This combination is invariant with respect to isometry transformation, and lifts up to a cohomology class. So locally there is a 3-form $\omega_3^0$ such that $d\omega_3^0 = {\rm Tr}\left[\mathcal R^2\right]$.
Note that making use of K\"ahler geometry, the bootstrapping procedure is similar to that of non-abelian gauge theories \cite{Zumino:1983rz}, if we consider the following relation:
\beq
\mathcal R_{i\bar j \,\,\,\,\,\,n}^{\,\,\,\,\,\,m}d\phi^i\wedge d\bar{\phi}^{\bar j} =- d\Gamma_{\,\,\,\,\,n}^{m} - \Gamma_{\,\,\,\,\,b}^{m}\wedge\Gamma_{\,\,\,\,\,n}^{b}\,,\quad \Gamma_{\,\,\,\,\,n}^{m} \equiv g^{m\bar a}\partial_i g_{n\bar a}d\phi^i\,.
\eeq
The isometry transformation is induced by the Killing vector field on the target manifold, which gives, 
\beq
\delta_{\epsilon^A} = dK_A+K_Ad\,,
\eeq
where the vector field is of the form \beq
K_A = K_A^i\partial_i + \bar K_A^{\bar i}\bar\partial_{\bar i}\,.
\eeq
Using the fact that the K\"ahler metric is compatible with the Killing vector fields, we get that
\beq
\delta_{\epsilon^A} \Gamma_{\,\,\,\,\,n}^{m} = -\partial (\partial_n K_A^m) -\partial_n K_A^a  \Gamma_{\,\,\,\,\,a}^{m} + \partial_a K_A^m \Gamma_{\,\,\,\,\,n}^{a}\equiv -d (\partial_m K_A^n) -[ \Gamma, \partial K_A]_{\,\,\,\,\,n}^{m}\,.
\eeq
Finally we have that $\delta_{\epsilon^A} \omega_3^0 = d \omega^1_2[K_A]$, which bears the form 
\beq
\omega_2^1\propto\left(\partial_m K_A^n d \Gamma_{\,\,\,\,\,n}^{m} 
-\bar\partial_{\bar n} \bar K_A^{\bar m} d\bar\Gamma_{\bar m}^{\,\,\,\,\,\bar n}
\right)\,.
\eeq
This, at the first order level, coincides with Eq.\,(\ref{nlanoexpr}).

\subsection{Global vs local anomalies from geometric point of view}
\label{4.2}

In this subsection we shall discuss technical points of the previous calculation, and then deduce the relation between  global and local anomalies. 

Indeed, we cheated a little in the previous calculation of the anomaly, and what has been hidden is the discussion on geometric condition for the anomaly. Notice that the Dirac operator $\mathcal D$, when restricted to the spin bundle on worldsheet, changes the helicity and hence maps $\psi_L$ to $\psi_R$. However it is not self-composable. This is because in our definition for $\mathcal D$, $\psi_L$ lives in the tangent space of a curved target space, while $\psi_R$ lives in a flat space. One consequence of this problem is that we actually do not have a precise definition for the functional determinant of $D$. This problem can be easily cured by choosing a local diffeomorphism $\mathcal{ E}[\phi, 0]: TX\to T\mathbb C^m$ and $\mathcal E^{-1}[0, \phi]: T\mathbb C^m\to TX$.\footnote{\,The diffeomorphisms $\mathcal E$ and $\mathcal E^{-1}$ are precisely the isomorphisms $T^{(+)}$ and $T^{(-)}$ used in Section 2 of \cite{Moore:1984ws}.} Note the parameter $0$ and $\phi$ in $\mathcal E$ merely indicates that $\psi_R$ is decoupled, and $\psi_L$ is coupled to $\phi$. Then we can compose $\mathcal D$ to obtain an elliptic operator whose image and source are the {\it same} Hilbert space:
\beq
\mathcal D^2:= \mathcal E \slashed {\mathcal D} P_L \mathcal E^{-1} \slashed{\partial} P_R + \mathcal E^{-1} \slashed {\partial} P_R \mathcal E  \slashed{\mathcal D} P_L\,.
\eeq
The functional determinant can now can be defined, and a regulator is introduced, by having that
\beq
 \delta_{\epsilon^A}\left({\rm det}i\mathcal D\right)[\phi, \bar\phi] = 
 {\rm lim}_{M\to \infty}  {\rm exp}\Big(\!- i\epsilon^A \!\!\int\! \dx\,{\rm Tr}\big[\mathfrak{Im}(\nabla_j K_A^i[\phi(x)]) \gamma_5 f\big({\mathcal D}^2\!/M^2\big)\big] \Big)
\eeq
for a smooth function $f(x)$ on $\mathbb R$ such that $f(0)=1$ and $f(\infty) = 0$.

Before proceeding to calculation, we immediately sense a problem, that the maps $\mathcal E$ and $\mathcal E^{-1}$ are only locally defined. Now if we want to patch the map to make it fibers nicely over the space of bosonic field $C^\infty(\Sigma, X)$ without ambiguity, we would have to view $\delta_{\epsilon^A}\left({\rm det} i\mathcal D\right)[\phi, \bar\phi]$ firstly as a complex line bundle, and impose the trivialization condition. In fact, one needs no worry here, if the aforementioned model is free of Moore-Nelson anomaly \cite{Moore:1984ws}. In their work, the condition to trivialize the line bundle $\left({\rm det}\,i \mathcal D\right)[\phi, \bar\phi]$ has been given. Suppose our model satisfies their condition, then $\left({\rm det}\,i \mathcal D\right)[\phi, \bar\phi]$ is a function of $\phi\in C^\infty(\Sigma, X)$, so is $\delta_{\epsilon^A}\left({\rm det}i \mathcal D\right)[\phi, \bar\phi]$. To conclude here, the vanishing of Moore-Nelson anomaly implies that there is no global obstruction for isometry.

Next, we shall look into the local anomaly. Now we consider the functional determinant $\left({\rm det}\,i\mathcal D\right)[\phi, \bar\phi]$ to be a function of $\phi\in C^\infty(\Sigma, X)$, then the variation of $\delta_{\epsilon^A}$ was induced by a vector field on $C^\infty(\Sigma, X)$. Locally we need that there exists a Lie algebroid structure on $TC^\infty(\Sigma, X)$ induced by the infinitesimal isometry on $X$, i.e., there is a subspace of $TC^\infty(\Sigma, X)$ over $C^\infty(\Sigma, X)$, which has a Lie bracket coming from the Lie brackets of Killing vector fields on $X$. This says that, the Lie algebra action can be realized on $\left({\rm det}\,i \mathcal D\right)[\phi, \bar\phi]$, i.e., $[\delta_{\epsilon^A}, \delta_{\epsilon^B}] = f^C_{AB}[\phi, \bar\phi]\delta_{\epsilon^C}$. Solving the Wess-Zumino consistency condition is equivalent to writing down the explicit form of the effective action (with  a
counterterm added) as a local functional. 

\subsection{The determinant line bundle analysis}
\label{4.3}

In previous section, we have seen that the relation between isometry anomaly, and the global anomalies for K\"ahlerian manifolds is the following. Once the global anomaly is absent, the functional determinant can be viewed as a function, as opposed to a section of a complex line bundle, over the space of bosonic field. Then, the isometry variation of the theory is via some selected vector fields acting on the determinant (ie, the effective action). The Wess-Zumino condition is then automatically satisfied. In the process of canceling the isometry variation, the counterterm is predicted indeed by the trivialization of a 4-form which represents the first Pontriyagin class. 

We want to clarify here when we have Hermitian vector bundles over a K\"ahler manifold, what do we mean by the first Pontryagin class. Indeed, the argument of \cite{Moore:1984ws} gave the anomaly in terms of a second real Chern character, which by definition is defined on real vector bundles by taking the complexification first, and then apply the complex Chern character. In this way, one verify that for real vector bundles, this second real Chern character precisely gives $p_1$ of the bundle, and in case of a complex vector bundle, this gives $2\, ch_2$ of the complex bundle.

Using Chern-Weil construction, choosing a connection $\Theta$ over the bundle $E$ one sees clearly that the 4-form representing the obstruction is 
$$\int_{Y\times \Sigma} ev^*{\rm tr}\Big[ \Big(\frac{i}{2\pi} \,\mathcal R^\Theta\Big)^{\!2} \Big]\,,$$
where $Y$ is an arbitrary 2-cycle in the space of bosonic field $C^\infty(\Sigma, X)$ and $ev$ is the evaluation map $$ev:\Sigma\times C^\infty(\Sigma, X)\to X\,.$$

We want to trivialize the expression, and one of the sufficient condition is that $ch_2$ vanishes before we pulling it back.
This Chern-Weil form of $ch_2$ can always be locally trivialized by the Chern-Simons transgression 3-form $CS(\Theta)$ on $X$. Moreover, if $ch_2$ is trivial, the Chern-Simons form is globally defined on $X$. Then we there is guarantee that the isometry variation of $CS(\Theta)$ is trivialized by a 2-form, which is able to compensate the anomalous transformation of the functional determinant\,, 
$$\delta_\alpha CS(\Theta) = d(\omega_2)\,, \omega_2 = {\rm tr}(\alpha d\Theta)\sim \delta_\alpha \Gamma_{\rm eff}\,.$$
So the counterterm in this case is given by $CS(\Theta)$.
If further more the Chern-Weil form turns out to be trivial, then $CS(\Theta)$ is a closed form, representing a cohomology class in $H^3(X;\mathbb Q)$. Then the counterterm to be added is genuinely 2-dimensional, which is determined by
\beq
\begin{split}
CS(\Theta) &=d\, \Omega_2\,,\\[1mm]
 \delta_\alpha CS(\Theta) &= d(\delta_\alpha\Omega_2)=d\, {\rm tr}(\alpha d\Theta)\,.
 \end{split}
 \eeq

Next we explain why holonomy anomaly, as arise genuinely from a gauge description, can be viewed as the nontriviality of certain determinant line bundle, the latter been discussed extensively by \cite{Moore:1984ws} and \cite{Freed}.

Starting with the bosonic field $g\in C^\infty(\Sigma, G)$ of the theory, we have that 
$$ev: C^\infty(\Sigma, G)\times \Sigma\to G\,,\quad (g,x)\mapsto g(x)\,,$$
and at the level of differential forms, we also have a pushforward map
$$e_*: \Omega^*(C^\infty(\Sigma, G)\times \Sigma)\to \Omega^*(C^\infty(\Sigma, G))$$
induced by integration along $\Sigma$.
The classical action of the theory should be viewed as a ${\rm dim }\Sigma$-form on $C^\infty(\Sigma, G)\times \Sigma$ pushed down to $\Omega^*(C^\infty(\Sigma, G))$, and hence is a function of the field. Path integral quantization amounts to say that there is also a certain pushforward map by integrating along $C^\infty(\Sigma, G)$. As we do not have applicable mathematical tools to rigidify the process, we shall just consider it as given by the canonical quantization. 

To build a coset model using chiral gauge method, we introduce a gauge field $A$ coming from a connection in $ Conn({\rm ad}_{\mathfrak h}P)$ for an adjoint $\mathfrak h$-bundle of \mbox{$H\!\to\! P\!\to\! \Sigma$}\,, and now the bosonic field $g$ is promoted to smooth sections \mbox{$g\!\in\! \Gamma(\Sigma, P\!\times_H\! G)$}. When the bundle $P$ has a global section, $g$ can be viewed as a $G$-valued smooth map from $\Sigma$. In the following analysis, we shall use a local trivialization of $P$ to write $g$ as a smooth map $U\to G$ for $U\subset \Sigma$ while keep in mind the nontrivial gluing of $g$ across open covers of $\Sigma$.

The infinite dimensional topological group $C^\infty(\Sigma, H)$ acts on the space of fields:
$$ C^\infty(\Sigma,H)\times \Gamma(\Sigma, P\times_H G)\to \Gamma(\Sigma, P\times_H G)): (h,g)\mapsto gh\,, $$
and
$$ C^\infty(\Sigma,H)\times Conn({\rm ad}_{\mathfrak h}P)\to Conn({\rm ad}_{\mathfrak h}P): (h, A)\mapsto  h^{-1}A h+h^{-1}d h\,.$$
The action is a functional over the space of field, which is invariant with respect to gauge transformation, and thus is a functional over the orbit space of diagonal action of gauge transformation, which we call the reduced space of field
$$\Gamma(\Sigma, P\times_H G)\times_{C^\infty(\Sigma,H)} Conn({\rm ad}_{\mathfrak h}P)\,.$$
Note that the gauge group acts on the bosonic field freely, so the quotient space can be taken as the honest orbit space without invoking ghost degree of freedom.

Now the gauge fixing is a local functional $f$ over the un-reduced space of field whose critical locus intersects $C^\infty(\Sigma,H)$-orbits transversely. By solving out the gauge fixing condition, one picks out a unique element in $\Gamma(\Sigma, P\times_H G)$ for each orbit, and correspondingly the action functional will be restricted to ${\rm Crit}(f) \times Conn({\rm ad}_{\mathfrak h}P)$, which models the reduced space of bosonic fields.
In the gauged formalism of bosonic homogeneous space sigma models, one fixes the gauge by asking the connection to be particular one pulled-back via $g$ from the principal $H$-bundle $\pi\!:\!G\to G/H$. Note that $G\to G/H$ is a Riemannian submersion, and hence this gives 
a subbundle $\pi^*T(G/H)\subset TG$. Now the Maurer-Cartan form splits into spin connection on $G/H$ and the vielbein 1-form
\beq
g^{-1}dg = \omega^i H_i+ e^a X_a\,.
\eeq

We now describe the fermions coming from $(0,1)$ supersymmetry. Those are, from the target side, sections of vector bundles associated to the principal $H$-bundle via the isotropic representation $\varrho$ where 
$ad\mathfrak g=ad\mathfrak h\oplus \varrho$ as before.

Let $S_L, S_R$ denote the bundles associated to $Spin(\Sigma)$ with half spin representation, and we have that 
$$\psi \in \Gamma(\Sigma, S_L\otimes g^* G\times_\varrho \mathfrak m)\,,\quad g\in \Gamma(\Sigma, P\times_H G)\,;$$
$$D_{RR}: \Gamma(\Sigma, S_L\otimes g^* G\times_\varrho \mathfrak m)\to \Gamma(\Sigma, S_R\otimes g^* G\times_\varrho \mathfrak m)\,. $$

There is also a linear gauge group action on fermions induced from the isotropic $H$-actions on $\varrho$. And due to the pull-backing of $g\in \Gamma(\Sigma, P\times_H G)$, the gauge connection $A$ is coupled to $\psi$. The Dirac operator we need to consider comes from a Dirac operator on the pulled-back bundle of $TG$ 
$$ D^{\varrho\oplus {\rm ad}\,\mathfrak h}_{RR}  = \partial_{RR} + g^{-1}\partial_{RR} g+ A_{RR}\,,$$
whose component in the isotropy representation is
\beq(D^{\varrho}_{RR})^{ab} = \partial_{RR}\delta^{ab} + \frac12 \,C_i^{ab}(A_{RR}^i+\omega^i_{RR})+ \frac 12 \,C_c^{ab}e^c_{RR}\,.\eeq

The operator is parameterized by $ \Gamma(\Sigma, P\times_H G) \times Conn({\rm ad}_{\mathfrak h}P) $ and is gauge covariant. If there is no chiral fermion anomaly, taking the functional determinant of it should result in a gauge invariant expression, and thus descending down to a functional over $\Gamma(\Sigma, P\times_H G)\times_{C^\infty(\Sigma,H)} Conn({\rm ad}_{\mathfrak h}P) $. The presence of fermionic anomaly is because of the fact that the fermionic effective action might be a section of a nontrivial complex line bundle over the space of fields for two reasons. Firstly, it is possible that the effective action is a line bundle already over the un-reduced total space $ \Gamma(\Sigma, P\times_H G) \times Conn({\rm ad}_{\mathfrak h}P) $ even before we check the gauge invariance; and secondly, it is possible that the nontriviality of the anomaly comes from the failure of descent condition at quantum level. 

Repeating the analysis in \cite{Moore:1984ws}, one knows that the line bundle is characterized by its first Chern class, which, upon integrating over a two-cycle in the space, gives the Chern number. In this way, one reduce the task of understanding the infinite dimensional space of field $\Gamma(\Sigma, P\times_H G)\times_{C^\infty(\Sigma,H)} Conn({\rm ad}_{\mathfrak h}P) $ to an arbitrary 2-dimensional 2-cycles in it.

We need to be more specific about the choice of 2-cycles. It is hard to lift up a 2-cycle from the base to the larger space precisely because the interaction between the 2-cycle in the base, and the gauge group. But here we have some convenient choice because of the special form of the Dirac operator. Note that the connection-dependence of the Dirac operator decouples into two parts
$$A_1=\frac12\left( \omega_{RR}- A_{RR}\right)\,,\quad A_2=A_{RR}|_\varrho +\frac12 \,e_{RR}\,,$$
the former is covariant with respect to gauge transformation, while the latter is not. In fact, $A_1$ is the difference of two connections on the very same bundle $P\to \Sigma$\,. 
This is based on two facts: 1) $\omega$ is a principal $H$-connection on $G\to G/H$; and 2) a section of the associated bundle $P\times_H G\to \Sigma$ can be used to pull the connection back to $P\to \Sigma$. To understand how the connection can be pulled back, it is enough to see that the sections pullback via
$ g\in \Gamma(\Sigma, P\times_H G) $, which is obvious. Along this line, one can view an element in $ \Gamma(\Sigma, S_L\otimes g^* G\times_\varrho \mathfrak m)$ as one in $ \Gamma(\Sigma, S_L\otimes P\times_\varrho \mathfrak m)$.
A characteristic computation at rational cohomology level would not depend on $A_1$. Now the analysis from determinant line bundle says that the anomaly is given by 
$$\int_{Y\times \Sigma} \hat{A}(Y\times \Sigma)\cdot ev^*ch(F^{A_2})\,.$$ 

The space $Y$ is a 2-cycle in the space of bosonic fields. On the one hand, if we ask Y to be a 2-sphere in $\Gamma(\Sigma, P\times_H G)$ which intersects gauge orbits transversely, then this expression gives rise to the known $p_1$ anomaly condition. If we take Y to be a 2-sphere suspended from gauge orbit \cite{AlvarezGaume:1983cs}, and use $A_{RR}$ as a representative for $A_2$, this gives the condition on non-abelian chiral gauge anomaly as shown in Sec.~\ref{3.1}.

\newpage

\section{\bf Conclusion}

In this paper we systematically study the anomalies in minimal $\mathcal N=(0, 1)$ and $(0, 2)$ supersymmetric sigma models on homogeneous spaces. The investigation starts from our previous obersvation \cite{CCSV2} on isometry/gauge anomalies correspondence for the sigma models realized in non-linear/linear gauged formalisms respectively. It leads us to consider more general holonomy anomalies and how to remove them. 

Following Polyakov and Wiegmann, we systematically explore the anomalous fermion effective action and obtain its explicit form. Later, in the procedure of mending the anomalous action, we derive an anomaly matching condition as criteria to sieve out ill-defined models. This condition is equivalent to the global topological constraint of $p_1(G/H)$ thoroughly  discussed by Moore and Nelson \cite{Moore:1984ws}. More importantly and surprisingly, we demonstrate that these local counterterms will further modify and constrain the behavior of the ``curable" theories in deep IR region. Supersymmetry will be broken in some theories, whereas some others flow to nontrivial infrared superconformal fixed points.  

In addition to the general discussion above, we also analyzed various concrete examples, applying the anomaly matching condition to different types of $G$ and $H$. We find that most survived minimal models are $\mathcal N=(0, 1)$ supersymmetric, while $\mathcal N=(0, 2)$ minimal models, due to their nontrivial center in $H$, are typically topologically obstructed. 

We also reveal an interesting correspondence between two-dimensional $\mathcal N\!=\!(0, 1)$ minimal sigma models and gauge theories,  analogous to  t'Hooft's anomaly matching observation in the four-dimensional case. Finally, we discussed the isometry/holo\-nomy anomalies and the anomaly matching condition from the standpoint of determinant line bundle. We obtained a more general expression on the anomaly equation with the help of a more powerful mathematical tool operative in fields configuration spaces.

Because of the simplicity of the  fermion sector in the minimal models we should expect that these models would be either destroyed or strongly constrained by anomalies. This expectation is more or less substantiated in this paper: our refined treatment of the anomalies and their remedies displays  very interesting features of the minimal $\mathcal N=(0, 2)$ and $(0, 1)$ sigma models.  Our subsequent work will continue along these lines. It should be interesting to work out some solid examples to further verify our results on the low-energy behavior of the minimal sigma models. Good candidates include models on $G/T^r$ (not necessarily maximal tori), since the complex structures
on them will enhance supersymmetry to $\mathcal N = (0, 2)$, which makes them particularly easy to handle. 

On the other hand, it is also noteworthy that the $\mathcal N=(0, 1)$ minimal sigma model on ${\rm SO}(2p)/({\rm SO}(p)\times {\rm SO}(p))$ corresponds to a $\mathcal N=(0, 1)$ two-dimensional gauge theory with the gauge group ${\rm SO}(p)$. It is, thus, interesting to ask whether or not every curable minimal model will have its corresponding gauge theory, and how to find them. Investigating these gauge theories may also shed light on the minimal sigma models, and \emph {vice versa}. We expect to answer some of these questions in the subsequent works.

\vspace{-4mm}

\section*{Acknowledgments}

X.C. thanks the Max-Planck Institute for Mathematics in Bonn for hospitality. The work of M.S. is supported in part by DOE grant DE-SC0011842. X.C. is supported by the Dorothea-Schl\"{o}zer Fellowship at the Georg-August 
Universit\"{a}t G\"{o}ttingen.

\end{document}